\documentclass[apj]{emulateapj}
\usepackage{apjfonts}
\newcommand{\ergcms}{erg cm$^{-2}$ s$^{-1}$}
\newcommand{\den}{N$_{\rm e}$}
\newcommand{\tem}{T$_{\rm e}$}
\newcommand{\cmthree}{cm$^{-3}$}
 
\newcommand{\chbeta}{C$_{H\beta}$}

\shorttitle{Abundances and Dust in Bulge PNe}
\shortauthors{Gutenkunst et al.}

\begin{document}

\title{Chemical Abundances and Dust in Planetary Nebulae in the
  Galactic Bulge}

\author{
{S.~Gutenkunst}\altaffilmark{1}, %
{J.~Bernard-Salas}\altaffilmark{1}, %
{S.~R.~Pottasch}\altaffilmark{2}, %
{G.~C.~Sloan}\altaffilmark{1}, and %
{J.~R.~Houck}\altaffilmark{1}
}

\email{sg283@cornell.edu}

\altaffiltext{1}{Center for Radiophysics and Space Research, Cornell
  University, 106 Space Sciences Building, Ithaca, NY 14853, USA.}

\altaffiltext{2}{Kapteyn Astronomical Institute, P.O. Box 800, NL 9700
  AV Groningen, the Netherlands.}

\begin{abstract}
  We present mid-infrared Spitzer spectra of eleven planetary nebulae
  in the Galactic Bulge.  We derive argon, neon, sulfur, and oxygen
  abundances for them using mainly infrared line fluxes combined with
  some optical line fluxes from the literature.  Due to the high
  extinction toward the Bulge, the infrared spectra allow us to
  determine abundances for certain elements more accurately that
  previously possible with optical data alone.  Abundances of argon
  and sulfur (and in most cases neon and oxygen) in planetary nebulae
  in the Bulge give the abundances of the interstellar medium at the
  time their progenitor stars formed; thus these abundances give
  information about the formation and evolution of the Bulge. The
  abundances of Bulge planetary nebulae tend to be slightly higher
  than those in the Disk on average, but they do not follow the trend
  of the Disk planetary nebulae, thus confirming the difference
  between Bulge and Disk evolution. Additionally, the Bulge planetary
  nebulae show peculiar dust properties compared to the Disk nebulae.
  Oxygen-rich dust feature (crystalline silicates) dominate the
  spectra of all of the Bulge planetary nebulae; such features are
  more scarce in Disk nebulae. Additionally, carbon-rich dust features
  (polycyclic aromatic hydrocarbons) appear in roughly half of the
  Bulge planetary nebulae in our sample, which is interesting in light
  of the fact that this dual chemistry is comparatively rare in the
  Milky Way as a whole.
\end{abstract}


\section{Introduction}

Abundances of planetary nebulae (PNe) have long been used to aid in
the understanding of the chemical history of the Milky Way.  Certain
elements such as argon and sulfur (and neon as long as the initial
mass is not near 3 $M_{\odot}$ and oxygen if initial mass of the
progenitor star is $\lesssim 5 M_{\odot}$, \citealt{karakas2003b,
  karakas2003a}) are not changed in the course of the evolution of the
low and intermediate mass precursor stars of PNe. Thus the abundances
of these elements give the chemical composition of the cloud from
which the PNe progenitor stars formed.  Many abundance studies have
been made of PNe (as well as stars and \ion{H}{2} regions) in the
Galactic Disk, leading to the determination of abundance gradients
across the Disk \citep[e.g.][]{shaver1983, rolleston2000,
  pottasch2006}.  However, due to the high extinction toward the
Bulge, there is a relative paucity of abundance studies of PNe as well
as stars and \ion{H}{2} regions in the Bulge.

Galactic bulges and spheroids may contain half of the stars in the
local universe \citep{ferreras2003}. Thus, understanding their
chemical evolution and formation is important to a general theory of
galaxy formation. Insights into our own Galactic Bulge formation have
implications for bulge formation in general.

Abundances of Galactic Bulge planetary nebulae (GBPNe) have the
potential to answer questions about how the Bulge formed.  For
example, what type of collapse formed the Bulge (dissipational or
dissipationless)? And, has secular evolution within the Galaxy since
Bulge formation caused a significant amount of star formation within
the Bulge \citep{minniti1995}?  At a bare minimum, a difference
between abundance gradients of PNe in the Bulge and Disk would imply
that they formed in separate processes.

The large extinction toward the GBPNe makes infrared (IR) lines
preferable to optical lines for determining their abundances.
Additionally, infrared lines provide essential data on important
ionization stages of argon, neon, and sulfur as well as \ion{O}{4}
for oxygen.  We complement the IR data with optical data where
necessary, so that we need no or only small ionization correction
factors (ICFs) to account for unobserved stages of ionization.
Finally, abundances derived from IR lines depend only weakly on the
electron temperature \citep{rubin1988, pottasch1999}.  All of these
factors lead to more accurately determined abundances than previously
possible with optical lines alone. Likewise, IR spectra are well
suited to study the various dust features of GBPNe because signatures
of both oxygen-rich dust (in the form of crystalline silicates) and
carbon-rich dust (in the form of polycyclic aromatic hydrocarbons;
PAHs) can be observed if they are present.

Abundances for a number of Galactic Disk planetary nebulae (GDPNe)
were determined with the use of spectra taken with the {\it Infrared
  Space Observatory} \citep[{\it ISO}; e.g.][]{pottasch2006}.
However, {\it ISO} lacked the sensitivity to study PNe further than
3--4 kpc away from the Sun.  As a result, {\it ISO} only studied two
Bulge PNe, M1-42 and M2-36.  Due to the better sensitivity of the
Infrared Spectrograph \citep[IRS;][]{houck2004} on the {\it Spitzer
  Space Telescope} \citep{werner2004} we are able to obtain spectra of
GBPNe closer to the Galactic Center; the furthest GBPN in our sample
is about 10 kpc from the Sun.

In this paper we present {\it Spitzer} IRS spectra of eleven
GBPNe. The next section describes the {\it Spitzer} data, while \S
\ref{supplementary_data} describes the supplementary data we use. In
\S \ref{data_analysis} we describe the data analysis, deriving ionic
and total abundances of argon, neon, sulfur, and oxygen. Additionally
we identify the crystalline silicate features and measure PAH
fluxes. Finally we discuss what our results imply for the evolution of
the Galactic Bulge and its PNe in \S \ref{discussion} and conclude in
\S \ref{conclusion}.

\section{{\it Spitzer} IRS Data}
\label{spitzer_obs_s}

\subsection{Observations}

We observed eleven GBPNe with the {\it Spitzer} IRS between September
2006 and September 2007 as part of the Guaranteed Time Observation
program 30550. In order to minimize slit losses, PeakUp with
0.4\arcsec\ positional accuracy was performed for the six PNe where it
was possible, while blind pointing with $\sim$1\arcsec\ positional
accuracy was done for the remaining five PNe.  We observed these PNe
with the IRS Short-Low (SL), Short-High (SH), and Long-High (LH)
modules, covering the wavelength range from 5 to 40 \micron.  In order
to subtract the background and minimize the effect of rogue pixels, we
took off-source observations for SH and LH; for the SL module we
subtracted the background by differencing the orders. The data were
taken in staring mode so that spectra were obtained at two nod
positions along each IRS slit. For SH and LH, a short exposure time of
six seconds was used to keep the bright lines from saturating, with a
total of four cycles for redundancy and to aid in the removal of
cosmic rays; for the SL module, data were taken in three cycles of
fourteen seconds each.  Table \ref{basicdata} gives the object names,
their Astronomical Observation Request (AOR) keys and coordinates.

\subsection{Source Selection}

The sources were selected to ensure they belong to the Bulge according
to the following criteria. (1) Foremost, the best criterion for
ensuring Bulge membership is having galactic coordinates $ |\ell| <
10\degr $ and $|b| < 10\degr$ \citep{pottasch1999}.  All of the
sources were selected to meet this criterion.  (2) We selected objects
with high radial velocities, except for two objects, PNG001.6-01.3 and
PNG002.1+03.3, where they are unknown and whose IRAS fluxes and
positions indicate that they are members of the Bulge,
\citep{acker1992}.  (3) Finally, the objects have diameters $\lesssim$
5\arcsec. \citet{pottasch1999} consider all PNe with diameters $>$
12\arcsec\ to be foreground objects, and thus choosing small diameters
helped to ensure Bulge membership. Table \ref{basicdata} gives the
radial velocities and diameters of our GBPNe.

Additionally, in order to make certain that we could get good {\it
  Spitzer} IRS spectra of the GBPNe, we chose isolated objects in the
IRAS Point Source Catalog (PSC) with small radial extent, accurate
coordinates, and observable intensities. While the IRAS PSC is not as
sensitive as our Spitzer observations (the IRAS PSC catalog is
sensitive to a couple hundred mJy whereas our Spitzer observations are
sensitive to a few mJy), we check that only one source is on the slit
during the data reduction.  The criterion of selecting PNe with small
sizes also ensured that nearly all of the flux from most of the PNe
could be observed within SL, the smallest IRS slit at 3.6\arcsec\
across. The sources also were chosen to have coordinates known to
better than 1.4\arcsec\ from the radio positions of
\citet{condon1998}, and these coordinates were refined with the 2MASS
catalog. Finally, we chose objects with radio fluxes at 21 cm (F$_{21
  cm}$) that implied IR fluxes bright enough (F$_{21 cm}$ $>$ 10 mJy)
to allow for short integration times, but dim enough (F$_{21 cm}$ $<$
50 mJy) to not saturate any of the IRS modules. Table \ref{basicdata}
gives the IRAS fluxes at 12 and 25 \micron\ as well as the radio
fluxes at 21 cm for our objects.

\subsection{Data Reduction}

\begin{figure*}
\epsscale{1.}
\plotone{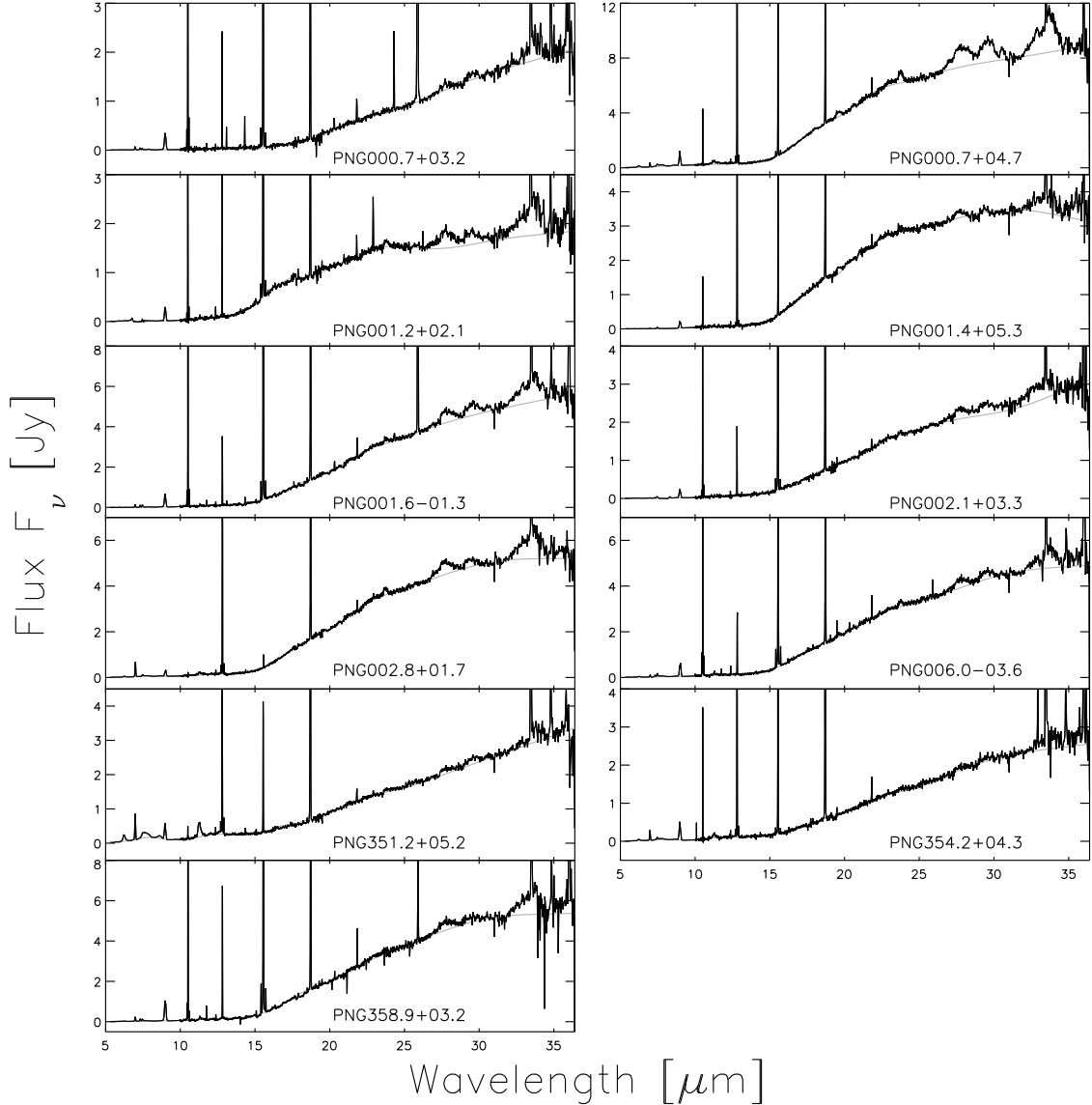}
\caption{Scaled, nod-averaged {\it Spitzer} IRS spectra (black) of our
  GBPNe. SL data is presented below 10 \micron, and HR data above 10
  \micron. The spline-fits to the continua (which we subtract from the
  spectra later in order to better view the crystalline silicate
  features) is over-plotted (grey). \label{silicates_oplot_cont}}
\end{figure*}

We start with basic calibrated data (bcd) from the {\it Spitzer
  Science Center's} pipeline version s15.3 or s16.1, and run it
through the IRSCLEAN\footnote{The IRSCLEAN program is available from
  the {\it Spitzer Science Center's} website at
  http://ssc.spitzer.caltech.edu} program to remove rogue pixels,
which uses a mask of rogue pixels from the same campaign as the data.
Then we take the mean of repeated observations (cycles) to improve the
signal to noise ratio.  After that the background is subtracted using
the off-source positions for SH and LH, and using the off-order for SL
(for example, SL1 nod1 - SL2 nod1). Next we use SMART
\citep{higdon2004} to manually extract the images, using full-slit
extraction for SH and LH and variable-column extraction for SL; we
also inspect the spectral profiles of each target with the Manual
Source Finder tool in SMART to ensure that only one source is within
the slit.  Spikes due to deviant pixels which the IRSCLEAN program
missed are removed manually in SMART. In order to account for flux
that fell outside of the IRS slits (due either to a slight mispointing
and/or the extended size of the GBPNe), we apply multiplicative
scaling factors to each order and nod.  The highest flux in LH sets
the scaling because LH is large enough to contain the entire flux of
all of our GBPNe.  Thus, one nod in LH is scaled to the other, the SH
nods are then scaled to LH, and the SL nods and orders are then scaled
to match SH.  Table \ref{scaling_factors} gives the scaling factors;
they are usually quite small ($\leq$1.20) except for three PNe where
the scaling factors in SL (the aperture with the smallest width) reach
up to 1.70.  Figure \ref{silicates_oplot_cont} plots the scaled and
nod-averaged spectra. We predict the 12 and 25 \micron\ IRAS fluxes
from these scaled IRS spectra and find generally good agreement with
the actual IRAS fluxes, confirming that only the IRAS source is within
the IRS slit.  Finally we use the gaussian profile fitting routine in
SMART to measure line fluxes for each nod position of the scaled
spectra. Table \ref{ir_lines} gives the observed nod-averaged line
fluxes. Uncertainties on the line fluxes are usually $\leq$10\%, with
uncertainties greater than this marked in the table.  A less-than sign
in Table \ref{ir_lines} indicates a 3$\sigma$ upper limit obtained
from the instrument resolution and the root mean square (RMS)
deviation in the spectrum at the wavelength of the line.

\begin{table*}
\setlength{\tabcolsep}{0.04in} 
\caption{Basic data for observed GBPNe \label{basicdata}}
\begin{tabular}{ccccccrcrcrclc}
\tableline
\tableline
\multicolumn{1}{c}{PNG} & \multicolumn{3}{c}{On Position\tablenotemark{a}} & \multicolumn{1}{c}{Off Position} & \multicolumn{1}{c}{log(F$_{H\beta}$)\tablenotemark{b}}  & \multicolumn{1}{c}{R$_{\odot,PN}$\tablenotemark{c}} & \multicolumn{1}{c}{R$_{GC}$\tablenotemark{d}} &  V$_{rad}$\tablenotemark{e} & Diam\tablenotemark{b} & \multicolumn{2}{c}{IRAS  Fluxes (Jy)\tablenotemark{f}} & \multicolumn{1}{c}{F$_{6 cm}$\tablenotemark{b}}& F$_{21cm}$\tablenotemark{g} \\
\cline{2-4} \cline{11-12} \\
    Number & AORkey   & RA          & DEC        & AORkey    &       (\ergcms)   & (kpc)~~&(kpc)~~~&(km s$^{-1}$)&(\arcsec)&F$_{12\mu m}$& F$_{25\mu m}$& ~~(mJy) & (mJy) \\ 
\tableline
000.7+03.2 & 17646848 & 17 34 54.71 & -26 35 56.9 & 17650176 & -13.40 $\pm$ 0.20 & 7.01  & 1.0$\pm$2.9  & -175 &  5.2   &$<$2.01&2.01  & 15    & 15.6  \\
000.7+04.7 & 17647616 & 17 29 25.97 & -25 49 07.1 & 17650432 & -13.90 $\pm$ 0.30 &  ...  & $<$4         & +40  &  2.7   &0.50   &6.56  & 27.7  & 12.8  \\
001.2+02.1 & 17648896 & 17 40 12.84 & -26 44 21.9 & 17650688 & -13.73 $\pm$ 0.10 & 6.64  & 1.4$\pm$2.8  & -172 &  4.0   &2.19   &3.00  & 26    & 24.2  \\
001.4+05.3 & 17647872 & 17 28 37.63 & -24 51 07.2 & 17650944 & -12.70 $\pm$ 0.30 & 7.90  & 0.2$\pm$1.9  & +42  &  5.0   &$<$0.28&2.71  & 13    & 13.8  \\
001.6-01.3 & 17649152 & 17 54 34.94 & -28 12 43.3 & 17651200 & -13.90 $\pm$ 0.30 &  ...  & $<$4         &  ... & 4.5    &$<$3.41& 3.49 &  ...  & 19.7  \\
002.1+03.3 & 17649408 & 17 37 51.14 & -25 20 45.2 & 17651456 &       ...         &  ...  & $<$4         &  ... & 4.8    &$<$1.93& 1.71 & ~5    & 46.0  \\
002.8+01.7 & 17649664 & 17 45 39.81 & -25 40 00.6 & 17651712 & -13.48 $\pm$ 0.10 & 7.50  & 0.6$\pm$2.5  & +164 &  3.8   &  ...  &  ... &  ...  & 13.8  \\
006.0-03.6 & 17648128 & 18 13 16.05 & -25 30 05.3 & 17651968 & -12.11 $\pm$ 0.02 & 4.91  & 3.2$\pm$2.1  & +136 &  5.1   &$<$1.45&3.35  & 51    & 41.2  \\ 
351.2+05.2 & 17648384 & 17 02 19.07 & -33 10 05.0 & 17652224 & -12.10 $\pm$ 0.10 & 7.69  & 1.2$\pm$1.2  & -128 &  5.0   &0.55   &1.70  & 12    & 14.4  \\
354.2+04.3 & 17648640 & 17 14 07.02 & -31 19 42.6 & 17652480 & -12.62 $\pm$ 0.10 & 10.71 & 2.8$\pm$4.0  & -75  &  4.0   &$<$0.34&1.40  & ~9.1  & 11.6  \\
358.9+03.2 & 17647104 & 17 30 43.82 & -28 04 06.8 & 17652736 & -13.03 $\pm$ 0.10 & 5.12  & 2.9$\pm$2.2  & +190 &  4.0   &$<$2.70&3.70  & 32    & 27.3  \\
\tableline
\end{tabular}

\tablenotetext{a}{RA and DEC are in J2000.0.  RA is in hours,
  minutes, seconds; DEC is in degrees, arcmin, arcsec.}

\tablenotetext{b}{From the Strasbourg-ESO Catalogue of Galactic
  Planetary Nebula \citep{acker1992}.  The diameter quoted here is the
  larger of the optical and radio diameters given in the catalogue.}

\tablenotetext{c}{Heliocentric distance, R$_{\odot,PN}$, from
  \citet{zhang1995}.}

\tablenotetext{d}{Galactocentric distance, R$_{GC}$, calculated
  assuming that the Sun is at 8.0 kpc from the Galactic Center. If
  R$_{\odot,PN}$ is unknown, then the PN is assumed to lie within 4
  kpc of the Galactic Center.}

\tablenotetext{e}{From \citet{durand1998} and \citet{beaulieu1999}.}

\tablenotetext{f}{From the IRAS catalogue of Point Sources, Version 2.0, \citet{helou1988}.}

\tablenotetext{g}{From \citet{condon1998}.}
\end{table*}

\begin{table*}
\begin{center}
\caption{Multiplicative scaling factors for GBPNe spectra \label{scaling_factors}}
\begin{tabular}{ccccccccccc}
\tableline
\tableline
PNG Number  &	LHn1 &	LHn2 &	SHn1 &	SHn2 &	SL1n1 &	SL1n2 &	SL3n1 &	SL3n2 &	SL2n1 &	SL2n2 \\
\tableline  
000.7+03.2  &   1.00 &	1.00 &	1.00 &	1.00 &	1.00 &	1.00 &	1.00 &	1.00 &	1.00 &	1.00 \\
000.7+04.7  &	1.02 &	1.00 &	1.00 &	1.00 &	1.00 &	1.00 &	1.00 &	1.00 &	1.00 &	1.00 \\
001.2+02.1  &	1.02 &	1.00 &	1.00 &	1.05 &	1.00 &	1.00 &	1.00 &	1.00 &	1.00 &	1.00 \\
001.4+05.3  &	1.03 &	1.00 &	1.15 &	1.17 &	1.15 &	1.15 &	1.00 &	1.15 &	1.00 &	1.00 \\
001.6-01.3  & 	1.05 &	1.00 &	1.15 &  1.15 &	1.15 &	1.15 &	1.50 &	1.50 &	1.50 &	1.50 \\	 
002.1+03.3  &	1.00 &	1.00 &	1.20 &	1.20 &	1.20 &	1.20 &	1.20 &	1.20 &	1.20 &	1.20 \\	
002.8+01.7  &	1.01 &	1.00 &	1.05 &	1.10 &	1.10 &	1.10 &	1.10 &	1.15 &	1.10 &	1.10 \\
006.0-03.6  &	1.02 &	1.00 &	1.15 &	1.15 &	1.15 &	1.15 &	1.60 &	1.70 &	1.50 &	1.40 \\
351.2+05.2  &	1.02 &	1.00 &	1.15 &	1.15 &	1.50 &	1.50 &	1.50 &	1.40 &  1.50 &  1.40 \\
354.2+04.3  &	1.02 &	1.00 &	1.15 &	1.15 &	1.20 &	1.20 &	1.20 &	1.20 &	1.20 &	1.20 \\
358.9+03.2  &	1.02 &	1.00 &	1.12 &	1.10 &	1.15 &	1.15 & 	1.00 &	1.20 &	1.00 &	1.00 \\	
\tableline
\end{tabular}
\tablecomments{LHn1 stands for LH nod 1, SL1n2 stands for SL order 1 nod 2, etc.}
\end{center}
\end{table*}


\begin{table*}
\setlength{\tabcolsep}{0.04in} 
\begin{center}
\caption{Observed infrared line fluxes \label{ir_lines}}
\begin{tabular}{lrrrrrrrrrrrr}
\tableline
\tableline
\multicolumn{1}{c}{Line} & \multicolumn{1}{c}{$\lambda$} &\multicolumn{11}{c}{Observed line fluxes for each object labeled by PNG number (10$^{-14}$ \ergcms)} \\
\cline{3-13}
                       & \micron~~  &  000.7+03.2                  & 000.7+04.7                  & 001.2+02.1                  & 001.4+05.3                  &    001.6-01.3            & 002.1+03.3                  & 002.8+01.7                  & 006.0-03.6                  & 351.2+05.2                  & 354.2+04.3                  & 358.9+03.2                  \\
\tableline
     $[$\ion{Ar}{2}$]$ &       6.99 &       26.15                  &     111.62                  &    $<$8.53                  &       6.63\tablenotemark{a} &   51.51\tablenotemark{a} &    $<$4.80                  &     253.42                  &      35.44\tablenotemark{a} &     304.57                  &      92.77                  &      85.93                  \\
      \ion{H}{1}(6-5)+ &       7.46 &       13.47                  &      36.22\tablenotemark{a} &      22.15                  &      19.40                  &   49.27                  &      18.57\tablenotemark{b} &      31.19\tablenotemark{a} &      61.70                  &      28.02\tablenotemark{a} &      19.67                  &      38.81                  \\
     $[$\ion{Ar}{5}$]$ &       7.90 &        8.82\tablenotemark{a} &   $<$13.74                  &    $<$2.95                  &    $<$2.26                  &$<$12.65                  &    $<$5.38                  &    $<$7.96                  &   $<$10.93                  &   $<$14.98                  &    $<$8.17                  &    $<$9.49                  \\
     $[$\ion{Ar}{3}$]$ &       8.99 &      162.55                  &     397.33                  &     124.24                  &      90.22                  &  278.81                  &     100.22                  &     108.59                  &     260.64                  &     173.77                  &     192.54                  &     472.17                  \\
     $[$\ion{S}{4}$]$ &      10.52 &     1401.40                  &     186.81\tablenotemark{a} &     573.91                  &      67.45                  & 1882.50                  &     688.40                  &      10.42\tablenotemark{b} &    2179.00                  &      19.77                  &     182.90                  &    1593.55\tablenotemark{a} \\
      \ion{H}{1}(7-6)+ &      12.37 &        5.35\tablenotemark{a} &      14.18\tablenotemark{a} &      10.53\tablenotemark{a} &       6.70\tablenotemark{a} &    9.37\tablenotemark{a} &       7.37\tablenotemark{b} &      11.16\tablenotemark{a} &      17.13                  &       8.10\tablenotemark{a} &       7.58\tablenotemark{b} &      11.89\tablenotemark{a} \\
     $[$\ion{Ne}{2}$]$ &      12.82 &       88.21                  &    1414.80                  &     133.51                  &     408.05                  &  142.14                  &      72.09                  &    1188.85                  &     106.01                  &    1132.05                  &     687.65                  &     228.45                  \\
     $[$\ion{Ar}{5}$]$ &      13.10 &       14.98\tablenotemark{a} &    $<$3.69                  &    $<$3.30                  &    $<$5.12                  &    9.25                  &    $<$4.47                  &    $<$4.73                  &    $<$4.35                  &    $<$3.90                  &    $<$4.25                  &    $<$3.75                  \\
     $[$\ion{Ne}{5}$]$ &      14.32 &       22.06                  &    $<$3.09                  &    $<$2.40                  &    $<$2.61                  &   11.74                  &    $<$2.63                  &    $<$3.63                  &    $<$2.85                  &    $<$4.62                  &    $<$3.51                  &    $<$3.52                  \\
     $[$\ion{Ne}{3}$]$ &      15.56 &     1590.25                  &    1466.85                  &    1500.55                  &     333.59                  & 3669.75                  &    1313.55                  &      17.82                  &    3455.80                  &     126.08                  &     676.02                  &    5245.25                  \\
      $[$\ion{S}{3}$]$ &      18.73 &      419.10                  &     503.54                  &     344.60                  &     333.46                  &  737.29                  &     255.92                  &     601.26                  &     669.36                  &     795.01                  &     600.36                  &     866.11                  \\
     $[$\ion{Ar}{3}$]$ &      21.84 &       11.03\tablenotemark{a} &      22.48                  &       8.10                  &       7.01\tablenotemark{a} &   22.43\tablenotemark{b} &       5.91\tablenotemark{b} &      12.44\tablenotemark{b} &      20.29\tablenotemark{c} &       9.27\tablenotemark{a} &      11.78                  &      34.77                  \\
     $[$\ion{Ne}{5}$]$ &      24.30 &       30.29                  &    $<$8.73                  &    $<$2.71                  &    $<$3.18                  &    8.63\tablenotemark{b} &    $<$2.40                  &    $<$4.78                  &    $<$4.50                  &    $<$2.55                  &    $<$3.06                  &    $<$8.33                  \\
      $[$\ion{O}{4}$]$ &      25.91 &     3580.55\tablenotemark{a} &    $<$8.83                  &    $<$2.63                  &    $<$3.83                  & 1313.85                  &    $<$2.42                  &    $<$4.52                  &      16.94\tablenotemark{a} &    $<$3.33                  &    $<$2.85                  &     101.99                  \\
      $[$\ion{S}{3}$]$ &      33.50 &      344.92                  &     110.49                  &     190.76                  &     164.33                  &  264.60                  &     158.36                  &     227.22                  &     214.29                  &     499.02                  &     417.67                  &     267.07                  \\
     $[$\ion{Ne}{3}$]$ &      36.03 &      181.76                  &     121.71                  &     150.23                  &   $<$36.43                  &  350.23                  &     110.85                  &   $<$25.80                  &     283.73                  &   $<$29.97                  &      70.92\tablenotemark{a} &     401.05                  \\
\tableline
\end{tabular}
\tablecomments{The \ion{H}{1} lines at 7.46 \micron\ and 12.37
  \micron\ both have contributions from more than one \ion{H}{1} line,
  as discussed in \S \ref{ext_cor_section}.  All line fluxes are from
  HR spectra except the lines with $\lambda$ $<$ 10 \micron ~which are
  from SL. A less-than sign indicates a three sigma upper limit.  All
  line flux uncertainties are $\leq$10\% unless otherwise indicated.}

\tablenotetext{a}{Uncertainty between 10 and 20\%.}
\tablenotetext{b}{Uncertainty between 20 and 50\%.}
\tablenotetext{c}{Uncertainty between 50 and 100\%.}
\end{center}
\end{table*}

\section{Supplementary Data}
\label{supplementary_data}

\begin{table*}
\setlength{\tabcolsep}{0.05in} 
\caption{Extinction corrected optical line fluxes \label{op_lines}}
\begin{tabular}{lrrrrrrrrrrr}
\tableline
\tableline
\multicolumn{1}{c}{Line} & \multicolumn{1}{c}{$\lambda$} &\multicolumn{10}{c}{Extinction corrected line fluxes relative to H$\beta$=100 for each object labeled by PNG number} \\
\cline{3-12}
                  &  \AA~~~   &  000.7+03.2 & 000.7+04.7 & 001.2+02.1 & 001.4+05.3 & 001.6-01.3  & 002.8+01.7 & 006.0-03.6 & 351.2+05.2 & 354.2+04.3 & 358.9+03.2 \\
\tableline
$[$\ion{O}{2}$]$  &      3727 &      114.0  &       ...  &       ...  &       ...  &       ...   &       ...  &      55.13 &      59.60 &     130.0  &       ...  \\
$[$\ion{Ne}{3}$]$ &      3869 &       69.6  &       ...  &      32.86 &       ...  &       ...   &       ...  &      95.19 &       ...  &       9.70 &       ...  \\
$[$\ion{O}{3}$]$  &      4363 &       12.4  &       2.81 &       ...  &       ...  &       ...   &       ...  &       6.68 &       ...  &       ...  &       ...  \\
$[$\ion{Ar}{4}$]$+\ion{He}{1} & 4712 & 7.00 &       ...  &       ...  &       ...  &       ...   &       ...  &       1.91 &       ...  &       ...  &       ...  \\
$[$\ion{Ar}{4}$]$ &      4740 &        4.61 &       ...  &       ...  &       ...  &       ...   &       ...  &       1.93 &       ...  &       ...  &       ...  \\
$[$\ion{O}{3}$]$  &      4959 &      279.1  &     123.99 &     231.9  &     118.2  &     354.2   &       ...  &     369.54 &       7.27 &      46.41 &     339.1  \\
$[$\ion{O}{3}$]$  &      5007 &      790.4  &     360.37 &     728.0  &     304.7  &    1003.8   &      22.65 &    1067.55 &      25.89 &     136.5  &     989.4  \\
$[$\ion{S}{3}$]$  &      6312 &        1.86 &       1.37 &       1.34 &       0.71 &       ...   &       ...  &       2.08 &       ...  &       1.01 &       2.02 \\
$[$\ion{S}{2}$]$  &      6717 &       13.35 &       2.76 &       2.77 &       2.70 &       6.40  &       4.77 &       4.35 &       6.46 &       5.96 &       6.68 \\
$[$\ion{S}{2}$]$  &      6731 &       19.47 &       5.12 &       4.37 &       3.30 &      14.38  &       9.06 &       7.83 &       9.31 &       9.36 &      12.41 \\
$[$\ion{Ar}{5}$]$ &      7005 &        0.67 &       ...  &       ...  &       ...  &       ...   &       ...  &       ...  &       ...  &       ...  &       ...  \\
$[$\ion{Ar}{3}$]$ &      7135 &       30.79 &      29.22 &      15.79 &      12.24 &      21.54  &       6.56 &      16.15 &       4.86 &       9.34 &      30.66 \\
$[$\ion{Ar}{4}$]$ &      7236 &        ...  &       0.96 &       ...  &       ...  &       ...   &       ...  &       ...  &       ...  &       ...  &       ...  \\
$[$\ion{Ar}{4}$]$ &      7264 &        ...  &       1.06 &       ...  &       ...  &       ...   &       ...  &       ...  &       ...  &       ...  &       ...  \\
$[$\ion{O}{2}$]$  &      7325 &        5.15 &      11.42 &       7.10 &       8.33 &       ...   &      3.87  &       6.29 &       0.91 &       2.07 &       7.73 \\
\tableline
\end{tabular}

\tablecomments{No optical line fluxes for PNG002.1+03.3 were found in
  the literature.}

\tablerefs{\citet{acker1991}, \citet{ratag1997},
  \citet{cuisinier2000}, \citet{escudero2004}, and \citet{wang2007}.}

\end{table*}


We supplement our IR data with optical and radio data from the
literature to aid abundance determinations for three reasons.  First,
we use the observed H$\beta$ and 6 cm radio fluxes to derive the
extinction toward GBPNe. Table \ref{basicdata} gives these fluxes from
the Strasbourg-ESO Catalogue of Galactic Planetary Nebula
\citep{acker1992}. Second, we adopt electron temperatures derived from
ratios of optical line fluxes (discussed in \S
\ref{Te_Ne_section}). Third, we use optical line fluxes for ions not
observable in our IR spectra (specifically lines fluxes of
\ion{Ar}{4}, \ion{S}{2}, \ion{O}{2}, and \ion{O}{3}) to reduce the
need for ICFs. (As an aside, no UV line data for any of the GBPNe in
our sample are available due to the large extinction toward the
Bulge.) When the optical line fluxes are given as observed line
fluxes, we apply the logarithmic extinction at H$\beta$ (\chbeta)
given in the paper to correct the lines for extinction because it
gives the correct Balmer decrement.  When more than one paper gives a
value for a line flux, we take the average (after correcting all line
fluxes for extinction), and Table \ref{op_lines} gives the optical
extinction corrected line fluxes adopted for the calculation of
abundances.


All of the PNe in this sample should be within $\sim$4 kpc or less of
the Galactic Center because they were selected to be members of the
Bulge.  In order to determine approximately where they are within the
Bulge to place them on a plot of abundance versus galactocentric
distance, we adopt the heliocentric statistical distances from
\citet{zhang1995}.  We chose these distances because \citet{zhang1995}
lists distances to more of our objects than other studies, such as
\citet{vandesteene1995} and \citet{cahn1992}.  An accurate statistical
distance scale for PNe is difficult to obtain, and controversies exist
as to which statistical distance scales are the best: for example
\citet{bensby2001} find that Zhang's distance scale is good, while
\citet{ciardullo1999} find that it is not as good.  However,
regardless of which statistical distances we adopt, the conclusions of
the paper will remain unchanged because all of the PNe in our sample
are constrained to be in the Bulge by other criteria and we include
the large uncertainties that go with these statistical distances in
the data analysis.

We adopt the distance from the Sun to the Galactic Center, R$_{\rm
  o}$, from \citet{reid1993} who determines the best estimate of this
distance by taking a weighted average of the various determinations of
R$_{\rm o}$ from different methods.  \citet{reid1993} finds that
R$_{\rm o} = 8.0 \pm 0.5$ kpc, and this value seems to agree with more
current estimates of this distance
\citep[e.g.][]{lopez-corredoira2000, eisenhauer2005,groenewegen2008}.
Galactocentric distances (R$_{GC}$) are then calculated assuming this
R$_{\rm o}$, and uncertainties on R$_{GC}$ are calculated using
standard error propagation and assuming an uncertainty of 40\% on the
heliocentric distance (\citealt{zhang1995} estimates the accuracy of
the PN distance scale as 35-50\% on average).  If the distance to a PN
is unknown, we assume it is within 4 kpc of the Galactic Center.
Table \ref{basicdata} lists the heliocentric and galactocentric
distances for each object.

\section{Data Analysis}
\label{data_analysis}

\subsection{Abundances}

Before determining abundances for this sample of GBPNe, we must first
calculate and then correct for extinction.  Additionally we adopt
electron temperatures (\tem) from the literature and then employ
infrared line ratios to derive the electron densities (\den).  After
that we use the values of the above quantities to obtain abundances
for each ion.  The following subsection discuss the details of the
calculations of extinction, the selection of \tem\ and \den, and
finally the derivation of ionic and total abundances.

\subsubsection{Extinction Correction}
\label{ext_cor_section}

We first calculate the reddening correction by comparing the H$\beta$
flux predicted from IR hydrogen recombination lines to the observed
H$\beta$ flux (see Table \ref{compare_chbeta}).  In order to predict
the H$\beta$ flux from the IR \ion{H}{1} lines, we adopt the
theoretical ratios of hydrogen recombination lines from
\citet{hummer1987} and assume case B recombination for a gas at \tem =
10$^4$ K and \den = 10$^3$ \cmthree. The \ion{H}{1}(7-6) line at
12.37 \micron ~and the \ion{H}{1}(11-8) line at 12.39 \micron ~are
blended in the SH spectrum, and theoretically the \ion{H}{1}(7-6) line
contributes 89\% of total line flux. Similarly, nearby lines of
\ion{H}{1}(6-5), \ion{H}{1}(17-8), \ion{H}{1}(8-6), and
\ion{H}{1}(11-7) contribute to the \ion{H}{1} line at 7.46 \micron,
with the \ion{H}{1}(6-5) flux theoretically contributing 74\% of the
total line flux.  The contributions of additional lines are removed
before calculating the predicted H$\beta$ flux from the
\ion{H}{1}(7-6) and \ion{H}{1}(6-5) IR lines.

Additionally we calculate extinction by comparing the H$\beta$ flux
predicted from the radio flux at 6cm to the observed H$\beta$ flux.
We assume that \tem=$10^4$ K (and thus t $\equiv$ \tem/$10^4$ K = 1),
He$^+$/H$^+$=0.09, and He$^{++}$/H$^+$=0.03 and use the following
formula from \citet{pottasch1984}:
\begin{displaymath}
 {F(H\beta)}_{6cm}^{predicted} = { S_{6cm} \over 2.82 \times 10^9 t^{0.53} (1 + {He^+ / H^+} + 3.7{He^{++} / H^+})} 
\end{displaymath}
where $2.82 \times 10^9$ converts units so that $S_{6 cm}$ is in Jy
and F(H$\beta$) is in \ergcms.  Table \ref{basicdata} gives the values
for $S_{6 cm}$ and F(H$\beta$) while Table \ref{compare_chbeta} gives
the calculated values of the extinction.

For the abundance calculations, in order to weight the extinctions
calculated from both of the above methods equally, we use
\begin{displaymath}
{C_{H\beta,final} = \frac{C_{H\beta,HI(7-6)}}4 + \frac{C_{H\beta,HI(6-5)}}4 + \frac{C_{H\beta,radio}}2}
\end{displaymath} 
when we have extinctions from both \ion{H}{1} lines and the radio,
otherwise we just take an average (see Table \ref{adopted_param} for
these adopted values). There is no H$\beta$ flux available for
PNG002.1+03.3 and thus we adopt an extinction to it from the average
of the other GBPNe extinctions.  Table \ref{compare_chbeta} gives the
extinction values derived here along with those from the literature.
In general there is very good agreement between the different methods.

We use the extinction law from \citet{fluks1994} and assume the
standard $R_V$ of the Milky Way of 3.1.  However, there is evidence
that interstellar extinction is steeper than this toward the Bulge,
e.g. \citet{walton1993a} find that $R_V$=2.3 and \citet{ruffle2004}
find $R_V$ = 2.0. Nevertheless, abundances determined from IR lines
are not greatly affected by this change in $R_V$: an $R_V$=2.0 usually
changes their abundances by $\lesssim$5\% (and at most 10\%) compared
to the usual $R_V$=3.1. Thus, because the previous optical studies to
which we compare assumed R$_V$=3.1, and because the IR lines are even
less affected by the choice of R$_V$, we assume R$_V$=3.1.

\begin{table*}
\begin{center}
\caption{Comparison of the derived \chbeta~ with the literature \label{compare_chbeta}}
\begin{tabular}{ccccccccccccc}
  \tableline
  \tableline
  \multicolumn{1}{c}{PNG}&\multicolumn{3}{c}{This work\tablenotemark{a}}&\multicolumn{1}{c}{ARKS91}&\multicolumn{2}{c}{RPDM97}&\multicolumn{1}{c}{CMKAS00}&\multicolumn{1}{c}{ECM04}&\multicolumn{2}{c}{WL07}&\multicolumn{2}{c}{TASK92}\\
  \cline{2-4}
  \cline{6-7}
  \cline{10-11}
  \cline{12-13}
  Number & \ion{H}{1}(7-6) &\ion{H}{1}(6-5) & 6 cm &Balmer &Balmer&6 cm&  Balmer& H$\alpha$/H$\beta$ & H$\alpha$/H$\beta$ &6 cm  & Balmer & 6 cm \\
  \tableline
  000.7+03.2 &  2.10 &  2.00 &  2.05 &	2.17 &	2.35 &  2.26 &	2.11 &	...   &	...  &	...   & 2.2  & 2.0  \\
  000.7+04.7 &  3.02 &  2.93 &  2.81 &	3.33 &	...  &  ...  &	...  &	2.88  &	...  &	...   & 3.3  & 2.8  \\
  001.2+02.1 &  2.72 &  2.55 &  2.62 &	2.71 &	...  &  ...  & 	...  &	2.40  &	...  &	...   &	2.7  & 2.6  \\
  001.4+05.3 &  1.50 &  1.46 &  1.28 &	1.25 &	...  &  ...  &	1.45 &	...   &	...  &	...   &	1.36 & 1.3  \\
  001.6-01.3 &  2.84 &  3.06 &   ... &	3.35 &	...  &  ...  &	...  &	...   &	...  &	...   &	3.4: & ...  \\
  002.1+03.3 &   ... &   ... &   ... &	...  &	...  &  ...  &	...  &	...   &	...  &	...   &	...  & ...  \\
  002.8+01.7 &  2.50 &  2.45 &   ... &	3.07 &	...  &  ...  &	...  &	...   &	...  &	...   &	3.1  & ...  \\
  006.0-03.6 &  1.31 &  1.37 &  1.29 &	...  &	1.43 &  1.35 &	...  &	...   &	1.41 &	1.18  &	1.30 & 1.32 \\
  351.2+05.2 &  0.98 &  1.02 &  0.65 &	1.11 &	0.985&  0.735&	...  &	...   &	...  &	...   &	1.14 & 0.68 \\
  354.2+04.3 &  1.47 &  1.39 &  1.05 &	1.69 &	1.78 &  ...  &	...  &	...   &	...  &	...   &	1.67 & 1.08 \\
  358.9+03.2 &  2.08 &  2.09 &  2.01 &	2.22 &	2.23 &  2.15 &	2.29 &	...   &	...  &	...   &	2.2  & 2.04 \\
  \tableline
\end{tabular}

\tablenotetext{a}{See \S \ref{ext_cor_section}.}

\tablerefs{ARKS91 = \citet{acker1991}, RPDM97 = \citet{ratag1997},
  CMKAS00 = \citet{cuisinier2000}, ECM04 = \citet{escudero2004}, WL07
  = \citet{wang2007}, and TASK92 = \citet{tylenda1992}.}

\end{center}
\end{table*}

\subsubsection{Electron Temperature and Density}
\label{Te_Ne_section}

In order to derive abundances, we adopt two electron temperatures
(\tem): T[\ion{N}{2}] for the low-ionization potential ions
(\ion{Ar}{2}, \ion{Ne}{2}, \ion{S}{2}, and \ion{O}{2}), and
T[\ion{O}{3}] for the high-ionization potential ions.  Table
\ref{Te_Ne} gives the electron temperatures from the literature. When
possible, our adopted T[\ion{N}{2}] and T[\ion{O}{3}] are an average
of the values from the literature.  If temperatures are not available
in the literature, we assume the average value from the other GBPNe in
the sample ($<$T[\ion{N}{2}]$>$ = 8100 K and $<$T[\ion{O}{3}]$>$ =
10700 K). Abundances from IR lines depend only weakly on the adopted
\tem, and thus our assumption does not strongly affect our abundances
--- especially those of argon, neon, and sulfur which are mostly
determined from IR lines.

We determine electron densities (\den) from IR line ratios of
\ion{S}{3}, \ion{Ne}{3}, \ion{Ar}{3}, \ion{Ar}{5}, and \ion{Ne}{5}
(see Table \ref{Te_Ne}).  The \ion{S}{3} line ratio gives the best
estimate of \den, and thus we adopt it for the abundance analysis.
Densities determined from the other line ratios are more uncertain
because they either rely on at least one line with a weak flux, or the
density is outside the range of what the line ratio can accurately
measure.  The adopted \den\ from the \ion{S}{3} ratios have an average
of 3500 \cmthree, and range from 1000 to 9200 \cmthree.

\begin{table*}
\begin{center}
\caption{Electron Temperatures and Densities\label{Te_Ne}}
\begin{tabular}{l|cc|rrrrr}
\tableline
\tableline
\multicolumn{1}{c}{PNG}&\multicolumn{2}{c}{Temperature T$_e$ (K)}&\multicolumn{5}{c}{Density N$_e$ (\cmthree)}  \\
\cline{2-8}
Number     & [\ion{O}{3}]                                   & [\ion{N}{2}]                                          & [\ion{S}{3}]          & [\ion{Ne}{3}]          & [\ion{Ar}{3}]          & [\ion{Ar}{5}]    & [\ion{Ne}{5}]  \\
\tableline

000.7+03.2 & 10200\tablenotemark{b}, 13300\tablenotemark{e} & 8000\tablenotemark{c}, 8400\tablenotemark{e}          &  1000\tablenotemark{g} &   low                   &    120\tablenotemark{g} &  370\tablenotemark{g} & low            \\
000.7+04.7 & 10200\tablenotemark{d}, 10200\tablenotemark{b} & 8500\tablenotemark{d}                                 &  9200                  &  3100\tablenotemark{g}  &  10000                  &  ...                  & ...            \\
001.2+02.1 & 10200\tablenotemark{b}                         & ...                                                   &  2100\tablenotemark{g} &   low                   &   2000\tablenotemark{g} &  ...                  & ...            \\
001.4+05.3 & 10200\tablenotemark{b}                         & 8300\tablenotemark{c}                                 &  2500\tablenotemark{g} &   ...                   &    low	                 &  ...                  & ...            \\
001.6-01.3 & ...                                            & ...                                                   &  4700                  &   low                   &    500\tablenotemark{g} &  ...                  & 3000\tablenotemark{g}\\
002.1+03.3 & ...                                            & ...                                                   &  1700\tablenotemark{g} &  2200\tablenotemark{g}  &   7500	                 &  ...                  & ...            \\
002.8+01.7 & 10200\tablenotemark{b}                         & ...                                                   & 3900                   &   ...                   &    low	                 &  ...                  & ...            \\
006.0-03.6 &  9800\tablenotemark{e}, 9840\tablenotemark{f}  & 9300\tablenotemark{e}, 11370\tablenotemark{f}         & 5000                   &  3600\tablenotemark{g}  &    low	                 &  ...                  & ...            \\
351.2+05.2 & 10200\tablenotemark{b}, 9300\tablenotemark{e}  & 7000:\tablenotemark{e}, 6000\tablenotemark{a}         &  1700\tablenotemark{g} &   ...                   &  14000                  &  ...                  & ...            \\
354.2+04.3 & 10200\tablenotemark{b}                         & 6600\tablenotemark{e}, 6400\tablenotemark{a}          &  1400\tablenotemark{g} &   low                   &   5400\tablenotemark{g} &  ...                  & ...            \\
358.9+03.2 & 10200\tablenotemark{b}, 7700\tablenotemark{e}, 10400\tablenotemark{a} & 8300\tablenotemark{e}, 8200\tablenotemark{a}, 8900\tablenotemark{c} &  5300 &7700 &    low                  &  ...                  & ...            \\
\tableline
\end{tabular}

\tablecomments{Electron temperatures were taken from the literature,
  with references given by the table notes. We derive electron
  densities from IR line ratios; `low' indicates that the density is
  lower than the theoretical ratio can measure.}

\tablenotetext{a}{\cite{acker1991}} \tablenotetext{b}{\cite{cahn1992}}
\tablenotetext{c}{\cite{cuisinier2000}}
\tablenotetext{d}{\cite{escudero2004}}
\tablenotetext{e}{\cite{ratag1997}} \tablenotetext{f}{\cite{wang2007}}

\tablenotetext{g}{The measured line ratio is at the low end (in the
  non-linear regime) of the densities that the theoretical line ratio
  can measure.}

\end{center}
\end{table*}

\subsubsection{Ionic and Total Abundances}
\label{ionic_tot_abun_sxn}



\begin{table}
\begin{center}
\caption{Adopted parameters for determining abundances \label{adopted_param}}
\begin{tabular}{lcccccc}
\tableline
\tableline
PNG     &F$_{H_\beta,predicted}$         &C$_{H_\beta}$  &\den       &T$[$\ion{O}{3}$]$ & T$[$\ion{N}{2}$]$   \\
Number  & (10$^{-14}$ \ergcms)       &             &(\cmthree) & (K)              &  (K)                \\
\tableline
   000.7+03.2 &  483 &       2.05 &      1000 &     11800 & 8200  \\
   000.7+04.7 & 1328 &       2.89 &      9200 &     10200 & 8500  \\
   001.2+02.1 &  902 &       2.62 &      2100 &     10200 & 8100\tablenotemark{b}  \\
   001.4+05.3 &  630 &       1.38 &      2500 &     10200 & 8300  \\
   001.6-01.3 & 1280 &       2.95 &      4700 &     10700\tablenotemark{b} & 8100\tablenotemark{b}  \\
   002.1+03.3 &  662\tablenotemark{a} &1.87\tablenotemark{a} &1700&10700\tablenotemark{b} & 8100\tablenotemark{b} \\
   002.8+01.7 & 1071 &       2.47 &      3900 &     10200 & 8100\tablenotemark{b}  \\
   006.0-03.6 & 1791 &       1.32 &      5000 &      9800 &10300  \\
   351.2+05.2 &  816 &       0.82 &      1700 &      9800 & 6500  \\
   354.2+04.3 &  674 &       1.23 &      1400 &     10200 & 6500  \\
   358.9+03.2 & 1211 &       2.04 &      5300 &     14200 & 8500  \\
\tableline
\end{tabular}
\tablenotetext{a}{No extinction nor H$\beta$ flux is given in
  literature for PNG002.1+03.3, so we cannot calculate the extinction
  from our data. We use the average extinction of the other GBPNe
  (\chbeta = 1.87) for the extinction toward PNG002.1+03.3.}
\tablenotetext{b}{When we could not find T[\ion{O}{3}] or
  T[\ion{N}{2}] in the literature, we adopted the average value from
  the other GBPNe in this sample.}
\end{center}
\end{table}

\begin{table*}
\begin{center}
\setlength{\tabcolsep}{0.04in} 
\caption{Ionic Abundances \label{ionic_abundances}}
\begin{tabular}{lccrrrrrrrrrrr}
\tableline
\tableline
\multicolumn{1}{c}{Ion}&\multicolumn{1}{c}{x \tablenotemark{a}}&\multicolumn{1}{c}{$\lambda$}&\multicolumn{10}{c}{Ionic abundances for each object labeled by PNG number} \\
\cline{4-14}
                               &       &            & 000.7+03.2  & 000.7+04.7  & 001.2+02.1  & 001.4+05.3  & 001.6-01.3  & 002.1+03.3  & 002.8+01.7  & 006.0-03.6  & 351.2+05.2  & 354.2+04.3  & 358.9+03.2  \\
\tableline
Ar$^{+} $  &    -7 &       6.99\tablenotemark{b} &        5.97 &        9.39 &     $<$1.07 &        1.14 &        4.63 &     $<$0.81 &       26.90 &        1.74 &       47.20 &       17.60 &        7.69 \\
Ar$^{+2}$  &    -7 &       8.99                  &       39.60 &       42.00 &       17.90 &       16.60 &       29.50 &       18.30 &       13.30 &       17.70 &       24.00 &       32.30 &       43.70 \\
Ar$^{+2}$  &    -7 &       21.8\tablenotemark{b} &       35.10 &       34.70 &       15.30 &       18.40 &       31.80 &       14.40 &       20.70 &       20.60 &       18.60 &       27.80 &       44.40 \\
Ar$^{+2}$  &    -7 &       7135                  &       20.70 &       25.70 &       13.90 &       10.80 &       17.20 &       ...   &        5.76 &       15.70 &        4.71 &        8.23 &       14.90 \\
Ar$^{+3}$ +&    -7 &       4712                  &       11.50 &       ...   &       ...   &       ...   &       ...   &       ...   &       ...   &        6.88 &       ...   &       ...   &       ...   \\
Ar$^{+3}$  &    -7 &       4740\tablenotemark{b} &        9.59 &       ...   &       ...   &       ...   &       ...   &       ...   &       ...   &        5.96 &       ...   &       ...   &       ...   \\
Ar$^{+4}$  &    -7 &       7.88                  &        0.76 &     $<$0.44 &     $<$0.14 &     $<$0.15 &     $<$0.42 &     $<$0.34 &     $<$0.32 &     $<$0.25 &     $<$0.77 &     $<$0.50 &     $<$0.31 \\
Ar$^{+4}$  &    -7 &       13.1\tablenotemark{b} &        0.82 &     $<$0.11 &     $<$0.10 &     $<$0.23 &        0.24 &     $<$0.19 &     $<$0.14 &     $<$0.08 &     $<$0.13 &     $<$0.17 &     $<$0.10 \\
Ne$^{+} $  &    -5 &       12.8\tablenotemark{b} &        3.14 &       18.80 &        2.63 &       10.80 &        2.01 &        1.89 &       19.70 &        0.83 &       25.60 &       19.10 &        3.20 \\
Ne$^{+2}$  &    -5 &       15.5\tablenotemark{b} &       21.80 &        8.43 &       11.70 &        3.59 &       20.90 &       13.50 &        0.12 &       13.70 &        1.04 &        6.67 &       28.20 \\
Ne$^{+2}$  &    -5 &       36.0\tablenotemark{b} &       27.70 &        8.84 &       13.10 &     $<$4.54 &       23.30 &       12.80 &     $<$2.00 &       13.80 &     $<$2.87 &        7.97 &       25.90 \\
Ne$^{+2}$  &    -5 &       3869                  &        3.93 &       ...   &        3.04 &       ...   &       ...   &       ...   &       ...   &       10.40 &       ...   &        0.90 &       ...   \\
Ne$^{+4}$  &    -7 &       14.3\tablenotemark{b} &        3.89 &     $<$0.26 &     $<$0.23 &     $<$0.35 &        0.87 &     $<$0.34 &     $<$0.31 &     $<$0.15 &     $<$0.46 &     $<$0.42 &     $<$0.28 \\
Ne$^{+4}$  &    -7 &       24.3\tablenotemark{b} &        5.22 &     $<$1.67 &     $<$0.31 &     $<$0.54 &        1.04 &     $<$0.34 &     $<$0.61 &     $<$0.40 &     $<$0.29 &     $<$0.39 &     $<$1.02 \\
S$^{+} $   &    -7 &       6717\tablenotemark{b} &       14.50 &        8.22 &        4.13 &        4.06 &       14.40 &       ...   &        9.66 &        4.97 &       18.20 &       15.40 &       13.90 \\
S$^{+} $   &    -7 &       6731\tablenotemark{b} &       18.20 &        7.86 &        4.55 &        3.31 &       18.60 &       ...   &       11.00 &        5.23 &       18.70 &       18.40 &       14.50 \\
S$^{+2}$   &    -7 &       18.7\tablenotemark{b} &       99.30 &       77.60 &       49.20 &       66.50 &       88.00 &       46.90 &       81.10 &       57.20 &      118.00 &      105.00 &       94.00 \\
S$^{+2}$   &    -7 &       33.4\tablenotemark{b} &       92.30 &       71.10 &       45.50 &       62.90 &       86.50 &       43.30 &       75.70 &       55.10 &      115.00 &       99.90 &       85.10 \\
S$^{+2}$   &    -7 &       6312                  &       25.30 &       29.90 &       30.50 &       16.00 &       ...   &       ...   &       ...   &       55.10 &       ...   &       23.10 &       14.70 \\
S$^{+3}$   &    -7 &       10.5\tablenotemark{b} &       73.90 &        5.99 &       18.80 &        2.88 &       50.80 &       28.00 &        0.31 &       38.10 &        0.61 &        6.76 &       38.90 \\
O$^{+} $   &    -6 &       3728\tablenotemark{b} &      128.00 &       ...   &       ...   &       ...   &       ...   &       ...   &       ...   &       31.20 &      279.00 &      582.00 &       ...   \\
O$^{+} $   &    -6 &       7327\tablenotemark{b} &      225.00 &      179.00 &      261.00 &      240.00 &       ...   &       ...   &      113.00 &       32.90 &      218.00 &      534.00 &      143.00 \\
O$^{+2}$   &    -6 &       4959\tablenotemark{b} &      174.00 &      117.00 &      218.00 &      111.00 &      289.00 &       ...   &       ...   &      401.00 &        7.88 &       43.70 &      130.00 \\
O$^{+2}$   &    -6 &       5007\tablenotemark{b} &      170.00 &      118.00 &      237.00 &       99.50 &      284.00 &       ...   &        7.36 &      401.00 &        9.72 &       44.50 &      131.00 \\
O$^{+3}$   &    -6 &       25.8\tablenotemark{b} &      157.00 &     $<$0.43 &     $<$0.08 &     $<$0.18 &       41.80 &     $<$0.09 &     $<$0.16 &        0.41 &     $<$0.10 &     $<$0.10 &        3.11 \\
\tableline
\end{tabular}

\tablenotetext{a}{To get abundances, multiply numbers in table by 10$^x$.}

\tablenotetext{b}{Ionic lines used to calculate total elemental
  abundances (\S \ref{ionic_tot_abun_sxn}).  }

\end{center}
\end{table*}

\begin{table*}
\begin{center}
\caption{Comparison of Total Elemental Abundances for Individual Bulge PNe \label{total_abundances}}
\begin{tabular}{l|cccc|cccc|ccc|cccc|cccc}
\tableline
\tableline
\multicolumn{1}{c}{PNG}&\multicolumn{4}{c}{This work}&\multicolumn{4}{c}{RPDM97\tablenotemark{a}}&\multicolumn{3}{c}{CMKAS00\tablenotemark{b}}&\multicolumn{4}{c}{ECM04\tablenotemark{c}}&\multicolumn{4}{c}{WL07\tablenotemark{d}}    \\
\cline{2-20}
~~~Number     &Ar    &Ne    &S     &O    &Ar    &Ne    &S    &O    &Ar    &S    &O     &Ar    &Ne    &S    &O    &Ar    &Ne    &S    &O     \\
\tableline
   000.7+03.2 &  5.2 & 3.7  & 1.9  & 5.1 & 4.8  & 0.56 & 0.62& 2.1 & 7.1  & 2.0 & 10.0 & ...  & ...  & ... & ... & ...  & ...  & ... & ...  \\
   000.7+04.7 &  5.4 & 2.7  & 0.88 & 3.0 & ...  & ...  & ... & ... & ...  & ... & ...  & 4.4  & ...  & 0.36& 1.9 & ...  & ...  & ... & ...  \\
   001.2+02.1 &  2.0 & 1.5  & 0.70 & 4.9 & ...  & ...  & ... & ... & ...  & ... & ...  & 1.7  & 0.35 & 0.46& 3.0 & ...  & ...  & ... & ...  \\
   001.4+05.3 &  2.5 & 1.4  & 0.71 & 3.5 & ...  & ...  & ... & ... & 1.9: & 0.56& 4.9  & ...  & ...  & ... & ... & ...  & ...  & ... & ...  \\
   001.6--01.3&  4.6 & 3.2  & 1.6  & 3.3 & ...  & ...  & ... & ... & ...  & ... & ...  & ...  & ...  & ... & ... & ...  & ...  & ... & ...  \\
   002.1+03.3 &  1.8 & 1.5  & 0.73 & ... & ...  & ...  & ... & ... & ...  & ... & ...  & ...  & ...  & ... & ... & ...  & ...  & ... & ...  \\
   002.8+01.7 &  5.3 & 2.0  & 0.89 & 1.2 & ...  & ...  & ... & ... & ...  & ... & ...  & ...  & ...  & ... & ... & ...  & ...  & ... & ...  \\
   006.0--03.6&  2.8 & 1.9  & 0.99 & 4.3 & 2.3  & 1.1  & 0.85& 4.8 & ...  & ... & ...  & ...  & ...  & ... & ... & 1.9  & 0.98 & 1.3 & 4.6  \\
   351.2+05.2 &  7.1 & 2.7  & 1.4  & 2.6 & 2.9  & ...  & 2.3 & 2.2 & ...  & ... & ...  & ...  & ...  & ... & ... & ...  & ...  & ... & ...  \\
   354.2+04.3 &  5.3 & 2.6  & 1.3  & 6.0 & 3.8  & 1.6  & 0.71& 7.2 & ...  & ... & ...  & ...  & ...  & ... & ... & ...  & ...  & ... & ...  \\
   358.9+03.2 &  6.5 & 4.0  & 1.4  & 2.8 & 5.6  & 5.1  & 0.36& 9.3 & 3.6: & 1.2 & 6.3  & ...  & ...  & ... & ... & ...  & ...  & ... & ...  \\
\tableline
\end{tabular}
\tablecomments{Obtain abundances relative to hydrogen by multiply the
  numbers in table by 10$^x$ where x is -6 for argon, -4 for neon, -5
  for sulfur, and -4 for oxygen.}  
\tablenotetext{a}{\citet{ratag1997}.}
\tablenotetext{b}{\citet{cuisinier2000}.  The colon (:) denotes low
  quality abundances due to a lack of data.}
\tablenotetext{c}{\citet{escudero2004}.}
\tablenotetext{d}{\citet{wang2007}.}
\end{center}
\end{table*}

\begin{figure*}
\epsscale{1.}
\plotone{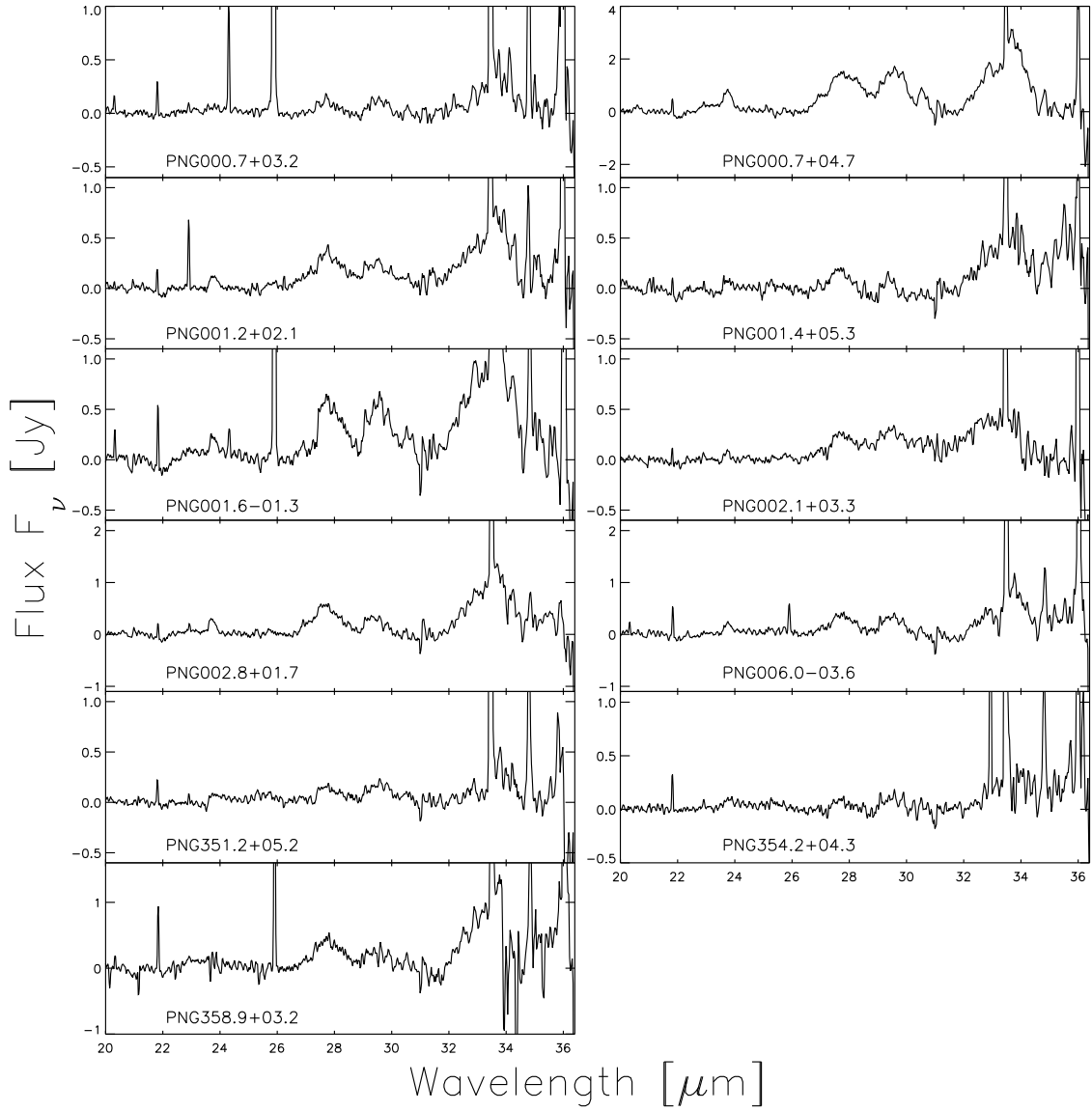}
\caption{Continuum subtracted spectra showing the crystalline silicate
  features. See Figure \ref{silicates_oplot_cont} for the spline fit
  to the continuum. \label{silicates_sub_cont}}
\end{figure*}


\begin{table*}
\begin{center}
\caption{Net PAH Feature Fluxes \label{pah_fluxes}}
\begin{tabular}{ccccccc}
\tableline
\tableline
\multicolumn{1}{c}{PNG} & \multicolumn{6}{c}{Net PAH Feature Fluxes ( x 10$^{-20}$ W cm$^{-2}$)} \\
\cline{2-7}
Number     & 6.2\micron~LR  & 7.7\micron~LR  & 8.6\micron~LR    & 11.2\micron~LR   & 11.2\micron~HR & 12.7\micron~HR \\
\tableline	
000.7+04.7 & 15.0 $\pm$ 0.2 & 27.3 $\pm$ 0.6 &  4.6 $\pm$ 0.2   & 11.3  $\pm$ 0.2  & 16.5 $\pm$ 0.8 & 12.$\pm$ 6.   \\
002.8+01.7 &  5.3 $\pm$ 0.1 &  8.4 $\pm$ 0.7 &  2.49 $\pm$ 0.08 &  6.08 $\pm$ 0.07 &  7.6 $\pm$ 0.5 &  9.$\pm$ 6.   \\
006.0-03.6 &  4.0 $\pm$ 0.2 & 11.0 $\pm$ 0.8 &  2.0 $\pm$ 0.1   &  4.4  $\pm$ 0.1  &  7.4 $\pm$ 0.5 &  4.1$\pm$ 0.5 \\
351.2+05.2 & 29.0 $\pm$ 0.9 & 40.8 $\pm$ 0.7 & 10.0 $\pm$ 0.1   & 22.0  $\pm$ 0.3  & 24.0 $\pm$ 0.5 & 11.$\pm$ 4.   \\
354.2+04.3 &  5.0 $\pm$ 0.2 &  8.1 $\pm$ 0.3 &  2.2 $\pm$ 0.1   &  6.32 $\pm$ 0.08 &  7.4 $\pm$ 0.5 &  7.$\pm$ 4.   \\
358.9+03.2 &  2.1 $\pm$ 0.3 &  4.7 $\pm$ 0.6 &  1.4 $\pm$ 0.2   &  3.4  $\pm$ 0.2  &  3.4 $\pm$ 0.8 &  1.$\pm$ 1.   \\
\tableline
\end{tabular}
\end{center}
\end{table*}

\begin{figure*}
\plotone{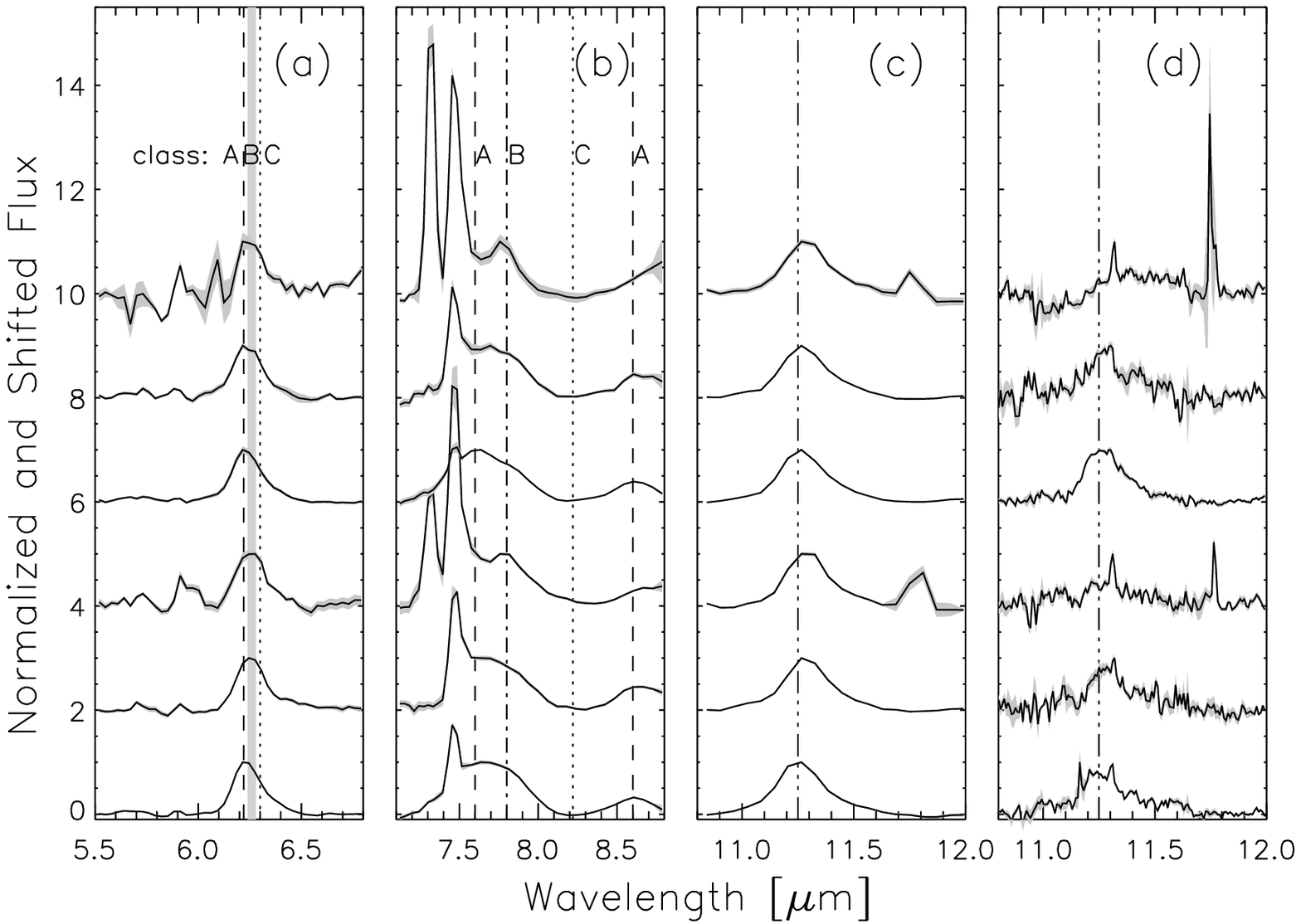}
\caption{Profiles of PAH features for PNG358.9+03.2, PNG354.2+04.3,
  PNG351.2+05.2, PNG006.0-03.6, PNG002.8+01.7, and PNG000.7+04.7 (from
  top to bottom).  {\bf (a)} Profiles of 6.2 \micron~ PAHs in LR. The dashed
  line indicates class A PAHs; the grey shaded area indicates class B;
  the dotted line indicates class C. {\bf (b)} Profiles of 7.7 \micron~ and
  8.6 \micron~ PAHs in LR. The dashed lines indicate class A PAHs; the
  dot-dashed line, class B; the dotted line, class C. {\bf (c)} Profiles of
  11 \micron~ PAH in LR. {\bf (d)} Profiles of 11 \micron~ PAH in
  HR. \label{pahs_figure}}
\end{figure*}

Table \ref{adopted_param} lists the parameters used in determining the
PNe ionic abundances.  Note that we use a predicted H$\beta$ flux from
the IR \ion{H}{1} lines in order to ensure that the hydrogen comes
from within the same slit as the IR forbidden lines.  Table
\ref{ionic_abundances} presents the ionic abundances themselves.  In
order to determine total elemental abundances, the ionic abundances
for each element are summed. When the ionic abundance can be
determined by more than one line, we choose the abundance(s) from the
most reliable line(s), and mark the lines used in Table
\ref{ionic_abundances}.  If necessary, the sum of the ionic abundances
for each element is then multiplied by an ionization correction factor
(ICF) to account for unobserved ions that are expected to be present.
For argon, we apply an ICF for the nine objects for which Ar$^{+3}$ is
not observed.  For neon, we apply an ICF for the four high ionization
nebulae with unobserved Ne$^{+3}$.  ICFs are generally small, and we
can derive accurate total elemental abundances for many objects,
especially for the elements of neon and sulfur whose abundances are
derived mainly from IR lines and which rarely need ICFs.  Table
\ref{total_abundances} presents the total elemental abundances.

The Ar$^{+3}$ abundance can only be determined directly from optical
lines for two of the GBPNe (PNG000.7+03.2 and PNG006.0-03.6). However,
the Ar$^{+3}$ abundance cannot contribute a large amount because the
abundance of Ar$^{+4}$ always accounts for $<$2\% of the total argon
abundance.  We adopt an ICF to account for unobserved Ar$^{+3}$
determined by Ar$^{+3}$ = 0.28*Ar$^{+2}$ because the two GBPNe with
observed Ar$^{+3}$ have Ar$^{+3}$/Ar$^{+2}$ = 0.27 and 0.29.
Additionally, the GDPNe in the sample of \citet{pottasch2006} for
which the ionic abundance of Ar$^{+4}$ is less than 2\% of the total
argon abundance (like our sample of GBPNe) have Ar$^{+3}$/Ar$^{+2}$
ranging from 0.15 to 0.68 with a mean of 0.30, so our assumption of
Ar$^{+3}$ = 0.28*Ar$^{+2}$ is justified.

The Ne$^{+3}$ abundance cannot be determined directly for any of our
GBPNe because its lines lie in the UV.  However, it is only expected
to contribute significantly if the O$^{+3}$ line is detected in the
IRS spectrum because its ionization potential (IP=63.45 eV) is near
that of O$^{+3}$ (IP = 54.93 eV).  The O$^{+3}$ line is detected
in only four of the spectra of our GBPNe (PNG000.7+03.2,
PNG001.6-01.3, PNG006.0-03.6, and PNG358.9+03.2). Thus, for these four
objects only, an ICF is necessary to account for unobserved
Ne$^{+3}$. Similarly to Ar$^{+3}$, the Ne$^{+3}$ cannot
contribute a huge amount because the abundance of Ne$^{+4}$ always
accounts for $<$1\% of the total abundance of neon.  Taking the
average of a sample of Galactic PNe, \citet{bernard-salas2008} find
that Ne$^{+3}$ = 0.35*(Ne$^{+2}$ + Ne$^{+4}$), and thus we adopt an
ICF determined by this to account for unobserved Ne$^{+3}$.

The uncertainties in the derived elemental abundances result from
uncertainties in the line fluxes, \chbeta, \tem, \den\ and ICFs.  The
measured \ion{H}{1} line fluxes typically have uncertainties
$\lesssim$20\%, while the measured fine structure line fluxes usually
have uncertainties $\lesssim$10\%.  Uncertainties are also introduced
into the line fluxes from the adopted scaling factors which are
typically $\lesssim$15\% for the high resolution lines (those lines
above 10 \micron), but reach 50--70\% for the low resolution lines for
three objects.  However, our scaling factors cannot be far off because
the \chbeta\ determined from \ion{H}{1} lines in in the high and low
resolution spectra agree well and the ionic Ar$^{+2}$ abundance
determined from \ion{Ar}{3} lines in the high and low resolution
spectra also agree well, even for the nebulae with the largest scaling
factors.  The uncertainties in \chbeta\ and \tem\ of $\sim$10\% each do
not have a large affect on the total elemental abundances of argon,
neon, and sulfur because these abundances are determined mainly from
IR lines; however, they will have a larger affect on the total
abundance of oxygen.  The uncertainty on \den\ is $\sim$30\%.  The
uncertainties on the ICFs for argon and neon are most likely less than
a factor of two, causing an abundance uncertainty for these elements
of $\lesssim$30\% due to the ICFs (when the ICFs are neccessary). A
comparison to optically derived abundances for the same objects by
various authors gives an estimate of the typical total abundance
uncertainty, which is $\sim$50\% \citep[e.g. this work,][]{gorny2004,
  bernard-salas2008}.

\subsection{Crystalline Silicates}

Crystalline silicate features are present around 28 and 33 \micron\ in
the spectra of all GBPNe in our sample, while no amorphous silicate
features are observed.  In order to illustrate the crystalline
silicate features more clearly, we define and subtract a continuum
determined by a smooth spline fit to feature-free regions of each
spectrum. Figure \ref{silicates_oplot_cont} shows the spline fit to
the spectral continua and Figure \ref{silicates_sub_cont} shows the
continuum-subtracted spectra.  The spline fit continuum is physically
meaningless, and we only use it to elucidate the crystalline silicate
features.  Following \citet{molster2000} we identify the 28 micron
complex (26.5 -- 31.5 \micron) and 33 micron complex (31.5 \micron\ to
past the end of our spectra) both as having features originating from
the magnesium-rich crystalline silicates forsterite (Mg$_2$SiO$_4$)
and enstatite (MgSiO$_3$).


The strength of the silicate emission bands can give an approximate
estimate of the crystalline dust temperatures.  \citet{matsuura2004}
note that the absence of a 23.7 \micron\ feature indicates that
forsterite is cooler than 100 K. This feature is either not present or
very weak in the spectra of our GBPNe, and thus the forsterite dust in
these objects must be cold, with a temperature $\lesssim$100 K.

\subsection{PAHs}

PAHs are present in six of the eleven GBPNe in our sample:
PNG000.7+04.7, PNG002.8+01.7, PNG006.0-03.6, PNG351.2+05.2,
PNG354.2+04.3, and PNG358.9+03.2. Absorption of energetic photons
excites the PAH emission features at 6.2, 7.7, 8.6, 11.2, and 12.7
\micron. PAHs which emit in this spectral range have on the order of
tens to hundreds of carbon atoms \citep{schutte1993}. C---C stretching
and bending or deformations causes the 6.2 and 7.7 \micron\ features,
while in-plane C---H bending produces the 8.6 \micron\ feature, and
out-of-plane C---H bending gives rise to the 11.2 and 12.7 \micron\
features \citep{allamandola1989}.

Table \ref{pah_fluxes} gives the net integrated PAH fluxes.  We
calculate these by first subtracting a spline-fit continuum and then
summing the remaining flux in each PAH wavelength range; if atomic
lines are present, we subsequently subtract their flux to arrive at
the net PAH flux.  For the 7.7 \micron\ PAH, we subtract the
\ion{H}{1} line at 7.46 \micron\ (and for PNG006.0-03.6 and
PNG358.9+03.2 the 7.32 \micron\ line as well). For the 12.7 \micron\
PAH, we remove the contribution from the Ne$^+$ line at 12.81
\micron; however, the 12.7 \micron\ PAH is much weaker than the 12.81
\micron\ Ne$^+$ line, and thus the net 12.7 \micron\ PAH flux is
very uncertain.  The 11.29 \micron\ \ion{H}{1} is weak and always near
the 3-$\sigma$ upper limit in our spectra; it contributes less than
5\% to the 11.2 \micron\ PAH flux (except for PNG006.0-03.6 and
PNG358.9+03.2 where it may contribute up to 20\%) and we do not remove
it.  Figure \ref{pahs_figure} shows plots of the continuum-subtracted,
normalized PAH profiles.

\clearpage

\section{Discussion}
\label{discussion}

\subsection{Elemental Abundances}
\subsubsection{Comparison of Abundances of Individual Objects with the Literature}
\label{compare_w_lit_indiv}

In Table \ref{total_abundances} we compare total elemental abundances
from this work with abundances from four papers in the literature:
\citet{ratag1997,cuisinier2000, escudero2004, wang2007}. All of these
studies derive total elemental abundances from collisionally excited
optical lines, and therefore their abundances are more dependent on
the adopted extinction and electron temperature than the current
study. A detailed comparison with these studies is hindered by the
fact that only one study \citep{wang2007} lists their ionic abundances
and ICFs. For individual objects, our total elemental abundances of
argon, neon, and sulfur tend to be higher than the optical abundances.
This is due in part to the fact that the IR derived abundances for
ions of Ar$^{+2}$, Ne$^{+2}$ and S$^{+2}$ always give a higher ionic
abundance than the optically derived abundances for the same ions. On
the other hand, the total elemental abundances of oxygen derived here
do not have such a systematic offset because the main contributors to
the total oxygen abundance, O$^+$ and O$^{+2}$, are determined from
optical line fluxes which are taken from the same literature sources
to which we compare abundances; PNG002.1+03.3 does not have an oxygen
abundance listed in Table \ref{total_abundances} because we could not
find any optical line fluxes for this object.

{\bf Argon} The values for the total argon abundance in this work are
systematically higher than the values given in the literature (except
in one case where the values are close). Several factors can lead to
this offset: {\bf (1)} In most cases, total elemental argon abundances
in this work and prior studies of the GBPNe in our sample must use an
ICF to account for unobserved Ar$^{+3}$; different ICFs will lead to
different total argon abundances.  When the 4740 \AA\ Ar$^{+3}$ line
is observed, our elemental argon abundance value agrees to within 30\%
of the values in the literature.  {\bf (2)} For the low excitation PNe
(PNG002.8+01.7, PNG351.2+05.2, and PNG354.2+04.3, Excitation Class,
EC$\sim$2--3), the IR data show that Ar$^+$ contributes
significantly to the total argon abundance, and thus optical studies
without observed Ar$^+$ must either use an ICF to account for it
or risk underestimating the total argon abundance. {\bf (3)} When we
derive the Ar$^{+2}$ ionic abundance from the IR lines and the
optical 7135 \AA\ line, we always get a value from the IR lines that
is higher than that from the optical line (often within 50\%, but
sometimes off by a factor of a few), which causes many of our IR
derived total argon abundances to be systematically higher than those
derived in the literature.  This may be due to the uncertainty in \tem\
when using optical lines to derive the Ar$^{+2}$ abundance:
lowering the electron temperature by 1000 to 2000 K significantly
increases the Ar$^{+2}$ ionic abundance derived from the optical
line (while only slightly increasing the Ar$^{+2}$ ionic abundance
derived from the IR lines), bringing the optical Ar$^{+2}$
abundances into good agreement with the IR Ar$^{+2}$ abundances in
most cases.

{\bf Neon} The values for the total neon abundance are systematically
higher in this study than in the literature (in all except one case
where the values are close). The factors that may cause this are: {\bf
  (1)} The IR data show that Ne$^+$ is the dominant contributor
to the total elemental neon abundance in roughly half of our GBPNe.
There is no Ne$^+$ line observable in the optical, and thus the
optical studies have not observed the most important ionization stage
of neon for these PNe. {\bf (2)} Lines of Ne$^{+3}$ lie in the UV
part of the spectrum, and thus our study and previous optical studies
must use an ICF to account for it in high ionization nebulae
(PNG000.7+03.2, PNG001.6-01.3, PNG006.0-03.6, and PNG358.9+03.2);
different assumed ICFs could account for part of the discrepancy for
these PNe.  {\bf (3)} When we derive the Ne$^{+2}$ ionic abundance
from the IR lines and the optical 3869 \AA\ line, we always get a
value from the IR lines that is higher than that from the optical.
Similarly to Ar$^{+2}$, this may be due (at least in part) to the
uncertainty in \tem\ having large affects on the optically derived
abundances.  Lowering the electron temperature by 1000 to 2000 K
increases the Ne$^{+2}$ ionic abundance derived from the optical
line (while only slightly changing the Ne$^{+2}$ ionic abundance
derived from the IR lines), bringing the optical and IR derived
Ne$^{+2}$ abundances into better agreement.

{\bf Sulfur} Most of the values for our total sulfur abundance are
higher than those given in the literature. This is due in part at
least to having derived a higher S$^{+2}$ abundance from the IR
lines as compared to the optical line.  The major contributors to the
total elemental sulfur abundance are S$^{+2}$ and S$^{+3}$, both
observed in our IR spectra. The optical S$^{+2}$ line at 6312 \AA\
is often weak and quite sensitive to \tem, and S$^{+3}$ is not
observed in the optical \citep{ratag1997}.  We use optical lines to
determine the abundance of S$^+$, but this is not a major
contributor to the total sulfur abundance.

{\bf Oxygen} Our values for the total oxygen abundance usually agree
to within a factor of two of those in the literature, and often within
50\%. For the one case where we can compare to the study in the
literature with published ionic abundances \citep{wang2007}, the ionic
abundance of O$^+$ is higher by 50\% in this work than in that
study, but the ionic abundance of O$^{+2}$ (the dominant ion) is
lower by 10\% than in that study, and the total elemental oxygen
abundances agree within 10\%. The IR data show that for one object
(PNG000.7+03.2), the O$^{+3}$ contributes significantly ($\sim$30\%)
to the total oxygen abundance, and thus optical studies must either
use an uncertain ICF or underestimate the total oxygen abundance in
this object.

Considering that we employ more observed stages of ionization than
purely optical studies and also that we derive ionic abundances for
the major contributors to the total elemental abundances for argon,
neon, and sulfur from IR lines (which are less sensitive to \chbeta\
and \tem\ than abundances from optical lines), our GBPNe abundances for
these elements are more accurate than previous studies.  Our GBPNe
abundance of oxygen, however, should be of similar accuracy to
previous optical studies because we must rely on optical lines for the
dominant ionization stages, but we make a slight improvement by
measuring or placing an upper limit on the abundance from the
O$^{+3}$ infrared line.

\subsubsection{Comparison of Mean Abundances with the Literature}
\label{compare_w_lit_mean}

We compare our mean Bulge abundance from the GBPNe to mean Bulge
abundances derived from other GBPNe abundance studies, red giant
stars, and \ion{H}{2} regions in Table \ref{compare_abundances}.  The
mean abundances of our GBPNe generally agree well with mean abundances
of GBPNe determined from the optical studies. The mean neon abundances
are the most discrepant, with ours being a factor of $\sim$2 higher
than those in the literature (reasons for such a discrepancy are given
in \S \ref{compare_w_lit_indiv}).  Our mean argon and sulfur
abundances are within the range of the previous studies, while our
mean oxygen abundance is only slightly lower.

\citet{cunha2006} derive abundances for seven red giant stars in the
Bulge, \citet{lecureur2007} forty-seven, and \citet{fulbright2007}
twenty-five.  \citet{cunha2006} derive oxygen abundances from lines of
OH vibrational-rotational molecular transitions observed in infrared
spectra, while \citet{lecureur2007} and \citet{fulbright2007} derive
oxygen abundances from the [\ion{O}{1}] line at 6300 \AA\ in optical
spectra.  The oxygen abundances of our GBPNe fall well within the
range of values for red giants from these studies, but the mean oxygen
abundance of the GBPNe is a factor of $\sim$2 lower than that of the
red giants.  However, given the uncertainties, small sample size, and
different methods used, there is a good agreement.

\citet{simpson1995} give abundances derived from IR lines for 18
\ion{H}{2} regions between 0 and 10 kpc from the Galactic Center,
while \citet{martin-hernandez2002} use {\it ISO} spectra to derive
abundances of 26 \ion{H}{2} regions between 0 and 14 kpc (distances
for both studies were redetermined so that R$_\odot$=8.0 kpc).  In
order to determine a mean \ion{H}{2} region Bulge abundance from these
studies, we take the mean of all \ion{H}{2} regions in each study
within 4 kpc of the Galactic Center.  The Bulge \ion{H}{2} region
abundances from these two studies generally agree well with our GBPNe
abundances, but the Bulge oxygen abundance of \citet{simpson1995} and
Bulge argon abundance of \citet{martin-hernandez2002} are a factor of
$\sim$2 higher.  There are only 5 objects in the central 4 kpc of
\citet{simpson1995} and only 3 in the central 4 kpc of
\citet{martin-hernandez2002} (and only 11 in our GBPNe sample), and
thus the small size of the samples may suggest that the mean does not
reflect a true average of the whole Bulge population. Our mean sulfur
abundance is the same as that of \citet{martin-hernandez2002}, but
over a factor of two smaller than that of
\citet{simpson1995}. Interestingly, while our mean GBPNe neon
abundance is a factor of $\sim$2 higher than previous GBPNe studies,
it agrees very well with the mean Bulge \ion{H}{2} region neon
abundances from these studies.

In order to compare abundances across the Disk as well as the Bulge of
the Galaxy, we supplement our abundances of GBPNe with those of GDPNe
that are derived from mainly IR lines in a similar way to the
abundances derived in this work.  They are mostly from
\citet{pottasch2006} who use chiefly {\it ISO} data (excluding the
strange low metallicity Hu 1-2), and complemented with abundances of
several GDPNe using mainly {\it Spitzer} data: NGC 2392 (Pottasch et
al., submitted), M1-42 \citep{pottasch2007}, and IC 2448
\citep{guiles2007}, and additionally abundances of one PN (NGC 3918)
that uses data from IRAS \citep{clegg1987}. In Table
\ref{HII_abundances} we compare mean abundances of PNe and \ion{H}{2}
regions with galactocentric distances in the range 0--4 kpc (Bulge),
4--8 kpc (Inner Disk) and beyond 8 kpc (Outer Disk).  The abundances
from PNe agree reasonably well with the abundances from \ion{H}{2}
regions derived by \citet{martin-hernandez2002}, but do not agree as
well with the abundances from \ion{H}{2} regions derived by
\citet{simpson1995}.  Ratios of abundances of the various
$\alpha$-elements to each other (for example, Ne/S, S/Ar, Ne/O) in
both PNe and \ion{H}{2} regions show flat behavior with galactocentric
distance (within the uncertainties), as expected for elements which
are thought to be made in the same processes in massive stars.

\begin{table}
\begin{center}
\caption{Comparison of Mean Bulge Abundances \label{compare_abundances}}
\begin{tabular}{lcccc}
\tableline
\tableline
~~~~~Study   & Ar/H & Ne/H & S/H  & O/H \\
\tableline
\multicolumn{5}{l}{PNe} \\
~~~~~Current & 4.4  & 2.5  & 1.1  & 3.7 \\
~~~~~RPDM97  & 3.8  & 0.98 & 1.0  & 5.2 \\
~~~~~CMKAS00 & 2.1  & ...  & 0.78 & 5.4 \\
~~~~~ECM04   & 4.7  & 0.75 & 0.63 & 3.9 \\
~~~~~WL07    & 2.0  & 1.2  & 1.1  & 5.1 \\
\tableline
\multicolumn{5}{l}{Red Giant Stars} \\
~~~~~CS06    & ...  & ...  & ...  & 7.3 \\
~~~~~LHZ07   & ...  & ...  & ...  & 8.8 \\
~~~~~FMR07   & ...  & ...  & ...  & 6.2 \\
\tableline
\multicolumn{5}{l}{\ion{H}{2} Regions} \\
~~~~~SCREH95 & ...  & 2.5  & 2.7  & 12  \\
~~~~~MHPM02  & 7.9  & 2.4  & 1.1  & ... \\
\tableline
\end{tabular}
\tablecomments{Obtain abundances relative to hydrogen by multiply the
  numbers in table by 10$^x$ where x is -6 for argon, -4 for neon, -5
  for sulfur, and -4 for oxygen.}

\tablerefs{RPDM97 = \citet{ratag1997}, CMKAS00 =
  \citet{cuisinier2000}, ECM04 = \citet{escudero2004}, WL07 =
  \citet{wang2007}, CS06 = \citet{cunha2006}, LHZ07 =
  \citet{lecureur2007}, FMR07 = \citet{fulbright2007}, SCREH95 =
  \citet{simpson1995}, and MHPM02 = \citet{martin-hernandez2002}.}
\end{center}
\end{table}

\begin{table}
\begin{center}
\caption{Abundances of PNe and \ion{H}{2} regions across the Galaxy\label{HII_abundances}}
\begin{tabular}{ccccc}
  \tableline
  \tableline
  \multicolumn{1}{c}{Distance range}&\multicolumn{1}{c}{Ar/H}&\multicolumn{1}{c}{Ne/H}&\multicolumn{1}{c}{S/H} &\multicolumn{1}{c}{O/H} \\
  (kpc) &         &           &          \\
  \tableline
  \multicolumn{5}{l}{PNe: This work + others (see \S \ref{compare_w_lit_mean})} \\
  \tableline
  0--4   & 4.6 & 2.7 & 1.2 & 4.5 \\
  4--8   & 4.3 & 1.9 & 1.2 & 5.0 \\
  8--... & 2.7 & 1.1 & 0.63& 4.2 \\
  \tableline
  \multicolumn{5}{l}{\ion{H}{2} Regions: \citet{simpson1995}} \\
  \tableline
  0--4   &...  & 2.5 & 2.7 & 12  \\
  4--8   &...  & 1.5 & 1.2 & 5.6 \\
  8--... &...  & 0.68& 0.76& 3.6 \\
  \tableline
  \multicolumn{5}{l}{\ion{H}{2} Regions: \citet{martin-hernandez2002}}  \\
  \tableline
  0--4   & 7.9 & 2.4 & 1.1 &...  \\
  4--8   & 4.7 & 2.2 & 0.89&...  \\
  8--... & 2.6 & 1.2 & 0.65&...  \\
  \tableline
\end{tabular}
\tablecomments{Obtain abundances relative to hydrogen by multiply the
  numbers in table by 10$^x$ where x is -6 for argon, -4 for neon, -5
  for sulfur, and -4 for oxygen.}

\end{center}
\end{table}

\subsubsection{Nature of the Bulge}

The absence or presence of an abundance gradient in the Bulge (and the
magnitude of the gradient if present) gives insight into how the Bulge
formed. If the Bulge has an abundance gradient, then it formed by
dissipational collapse, where self-enhancement of abundances occurred
as the collapse continued inwards. However, if the Bulge does not have
an abundance gradient, then it formed by dissipationless collapse,
where mergers of small protogalactic pieces caused an inhomogeneous
collapse over a long period of time and the mergers mixed stars of
different ages and metallicities.  If the Bulge has only a shallow
abundance gradient, then the gravitational potential of the bar in our
Galaxy caused concentrated star formation at its center and the stars
eventually left the Disk to become (part of) the Bulge
\citep{minniti1995}.

Several (mainly optical) studies of GBPNe point toward a slightly more
metal-rich Bulge than Disk \citep{ratag1992,cuisinier2000,gorny2004,
  wang2007}.  However, \citet{ratag1992} find that the average
abundances of GBPNe cannot be predicted by the abundance gradient
observed for GDPNe, hinting that stars in the Bulge are a distinct
population from the Disk. Additionally, \citet{gorny2004} find that
the O/H gradient becomes shallower and may even decrease in the most
inner parts of the Disk based on their sample of GDPNe towards the
Galactic Center. On the other hand, \citet{exter2004} find essentially
no difference in abundances between their Bulge and Disk PNe samples;
however, their results also point to a discontinuation of the Disk
metallicity gradient. The large extinction toward the Bulge hinders
optical studies of GBPNe. Thus, in this work we seek to confirm the
results of the optical studies using mainly infrared data.

\begin{figure*}
\epsscale{1.}
\plotone{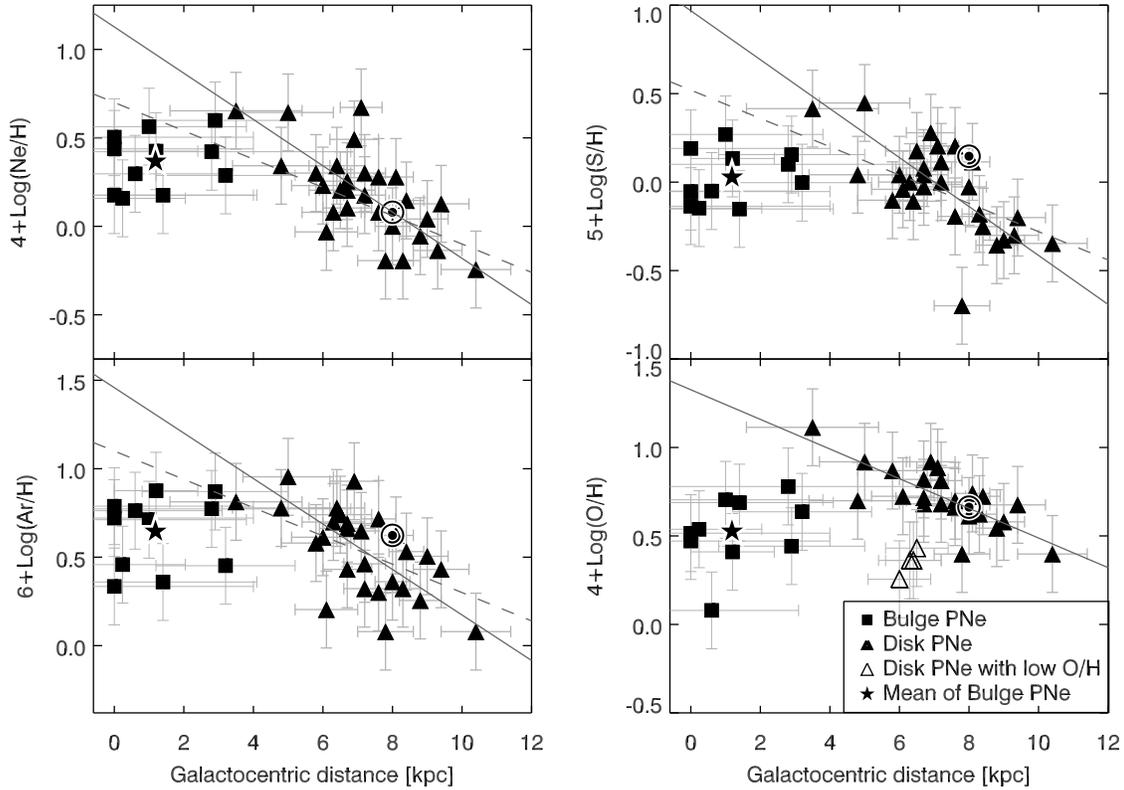}
\caption{The argon, neon, sulfur, and oxygen abundances of GBPNe and
  GDPNe versus the galactocentric distance.  GBPNe are represented by
  filled squares and GDPNe by filled triangles while the assumed solar
  metallicity as discussed in \citet{pottasch2006} is represented by
  the sun symbol, and the star indicates the mean of the GBPNe
  abundances. The solid grey lines represent least squares fits to the
  GDPNe abundances, excluding from the fit to the oxygen abundance the
  four GDPNe that are thought to have depleted oxygen (as discussed in
  \citet{pottasch2006}; open triangles). The dashed grey lines in the
  plots of argon, neon, and sulfur represent the oxygen abundance
  slope passing through the solid line fit at 8 kpc. Coefficients of
  the linear fits to both GBPNe and GDPNe are given in Table
  \ref{fits_to_abund_grad}.  The GBPNe placed on the graph exactly at
  R$_{GC}$=0.0 kpc have unknown galactocentric distances and are not
  included when we perform a linear fit to the data. The y-axis range
  for all the plots spans 2.0 dex, so that equal slopes will look
  equal in the plots.  Distance uncertainties are propagated from the
  statistical distance uncertainties while abundance uncertainties are
  assumed to be 50\%. \label{plot_abun_vs_dist}}
\end{figure*}

In order to discover if the abundance trend in the Disk continues in
the Bulge, Figure \ref{plot_abun_vs_dist} shows abundances of argon,
neon, sulfur, and oxygen versus galactocentric distance for both the
GBPNe and GDPNe (GDPNe data discussed in \S \ref{compare_w_lit_mean}).
We fit lines to the plots of GBPNe and GDPNe abundances versus
galactocentric distance in this figure separately (excluding from the
fit to the oxygen abundance the four GDPNe that are thought to have
depleted oxygen due to hot bottom burning, as discussed in
\citealt{pottasch2006}), and Table \ref{fits_to_abund_grad} gives
parameters for these fits.  The elemental abundance gradients of the
GDPNe range from $-$0.08 to $-$0.14 dex/kpc and have uncertainties of
0.03--0.04 dex/kpc.  In Figure \ref{plot_abun_vs_dist} we also over
plot the oxygen abundance gradient passing through the fit to the
GDPNe abundances at 8 kpc on the plots for the other elements in order
to illustrate that the abundances of the GBPNe are not consistent with
the abundance versus galactocentric radius trend of GDPNe, whether the
abundance data is fit directly to determine the gradient or if the
shallower oxygen abundance gradient is assumed.  The GBPNe have
abundances significantly lower than the abundance in the Bulge
predicted by the GDPNe abundance gradients.

Unfortunately the uncertainties in our fit to the abundance gradient
of GBPNe do not allow us to determine if an abundance gradient is
present in the Bulge; thus we cannot conclude anything about the
specific method of Bulge formation. The large velocities of objects in
the Bulge may smear out any abundance gradient that was originally
present.  However, the GBPNe abundances clearly do not follow the
abundance gradient trend of GDPNe (see Figure
\ref{plot_abun_vs_dist}): while the GBPNe have slightly higher average
abundances that the GDPNe, they still fall far below the GDPNe
abundance gradient extrapolated into the Bulge.  This corroborates
optical studies which had previously shown a discontinuity between the
Bulge and Disk abundance gradients, confirming the distinct nature of
the Bulge compared to the Disk.

\begin{table}
\begin{center}
\caption{Parameters of linear fits to abundance gradients in GBPNe and GDPNe. \label{fits_to_abund_grad}}
\begin{tabular}{lcccc}
\tableline
\tableline
\multicolumn{1}{c}{Element}&\multicolumn{2}{c}{GBPNe}&\multicolumn{2}{c}{GDPNe} \\
\cline{2-3}
\cline{4-5}
          &     y-int (dex)  & slope (dex/kpc) &   y-int (dex)   &  slope (dex/kpc) \\
\tableline
    neon  &  0.0  $\pm$ 0.9  &  0.2  $\pm$ 0.3 &  1.1 $\pm$ 0.3  & -0.13 $\pm$ 0.04 \\
  sulfur  & -0.2  $\pm$ 0.6  &  0.2  $\pm$ 0.3 &  1.0 $\pm$ 0.3  & -0.14 $\pm$ 0.04 \\
   argon  & -12~  $\pm$ 13~  &  8~~  $\pm$ 260~&  1.5 $\pm$ 0.3  & -0.13 $\pm$ 0.04 \\
  oxygen  & -0.0  $\pm$ 0.9  &  0.3  $\pm$ 0.6 &  1.3 $\pm$ 0.3  & -0.08 $\pm$ 0.03 \\
\tableline
\end{tabular}
\end{center}
\end{table}

\subsection{Crystalline Silicates} 
\label{crystalline_silicates_discussion}

Prior to {\it ISO}, crystalline silicates had only been observed in
solar system comets \citep[e.g.][]{hanner1994} and in $\beta$ Pic, a
debris disk system \citep{knacke1993}.  {\it ISO} and now {\it
  Spitzer} have observed crystalline silicates in many
sources. However, it is remarkable that we observe crystalline
silicates in every single one of the GBPNe. We suggest here that this
is because the GBPNe have disks.

\begin{figure*}
\begin{flushleft}
\epsscale{1.}
\plotone{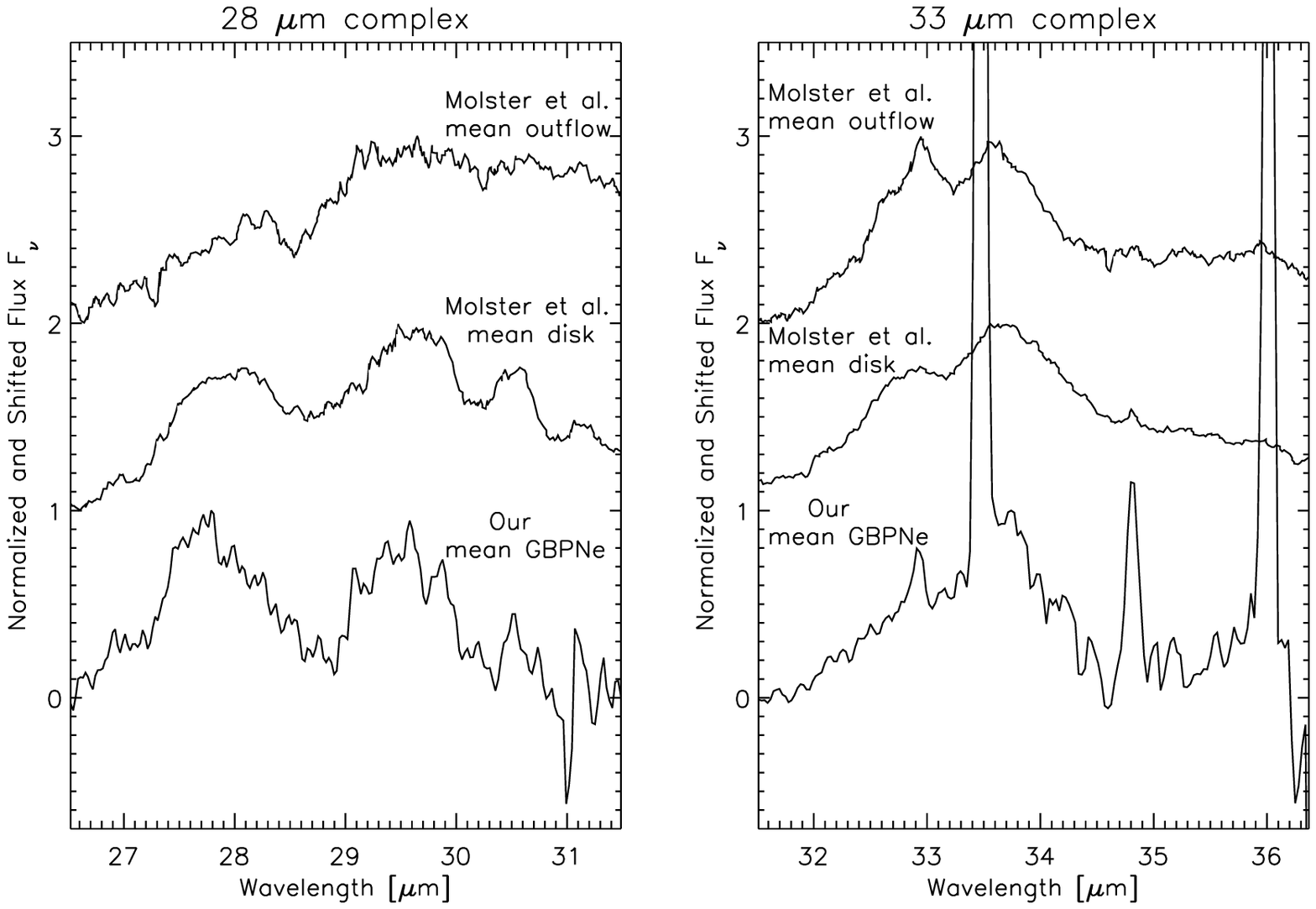}
\caption{Normalized and shifted mean 28 and 33 \micron\ spectra of our
  GBPNe compared to the mean disk and outflow sources from
  \citet{molster2002}.\label{mean_silicates}}
\end{flushleft}
\end{figure*}

In their {\it ISO} study of crystalline silicate dust around evolved
stars, \citet{molster2000} and \citet{molster2002} make mean continuum
subtracted spectra for sources which are thought to have a dusty disk
(disk sources) and sources which are expected to have a normal outflow
(outflow sources). They find that the dust features of disk and
outflow sources show definitive differences in strength, shape, and
position of their dust features.  In Figure \ref{mean_silicates} we
plot normalized mean spectra of our GBPNe for the 28 and 33 \micron\
features and compare them to the normalized mean disk and outflow
spectra from \citet{molster2002}.  Both the 28 and 33 \micron\
complexes in our GBPNe look similar to the mean disk sources in
\citet{molster2002}, but \citet{molster2002} have several cautions
about their mean spectra (for example, the {\it ISO} SWS band 3E,
which covers the $\sim$27.5--29.2 \micron, is known to have less
reliable calibration). However, the similarity of the crystalline
silicate dust features in our GBPNe to those of Molster's disk sources
gives indirect evidence that the silicates in our GBPNe are in disks.


If the crystalline silicates in these GBPNe are in fact in disks, then
they point toward binary evolution of the progenitor stars.
\citet{edgar2007} ran numerical models that show how a binary
companion can shape the AGB wind to form a crystalline dust torus.  In
their models, the shock temperatures reached when the wind blows past
the companion anneal the dust and make it crystalline. They conclude
that ``Crystalline dust torii provide strong evidence for binary
interactions in AGB winds.''  As we discuss later in \S
\ref{dual_chemistry}, over half of the GBPNe in this study show dual
chemistry, which also implies binary evolution.


In our GBPNe sample, all of the nebulae show crystalline silicates,
indicative of oxygen-rich material. Previous studies have found a low
C/O ratio in GBPNe compared to GDPNe
\citep[e.g.][]{walton1993a,wang2007,casassus2001}.  The higher
fraction of O-rich PNe in the Bulge compared to the Disk implies that
the Bulge should have a larger injection of silicate grains into its
interstellar medium (ISM) than the Disk \citep{casassus2001}.


\subsection{PAHs}

PAHs can be separated into different classes based on the position of
their 6.2 \micron\ and 7.7 \micron\ peaks.  Class A PAHs peak at
shorter wavelengths than class B, which peak at shorter wavelengths
than class C \citep{peeters2002}.  Figure \ref{pahs_figure} shows the
peak positions for the different classes of PAHs along with the GBPNe
PAH features.  The GBPNe in this study have class A, AB, and B PAHs,
and thus have PAH profiles similar to GDPNe.  The lack of type C PAHs
in the PNe indicates that their PAHs are all processed, i.e. the
aliphatic component is negligible \citep{sloan2007}.

The PAH flux ratios F$_{7.7 \micron}$/F$_{11.2 \micron}$ and F$_{6.2
  \micron}$/F$_{11.2 \micron}$ both trace the ionization fraction of
the PAHs, and are often plotted against each other in a figure.  The
GBPNe studied here have F$_{7.7 \micron}$/F$_{11.2 \micron}$ between 1
and 3, and F$_{6.2 \micron}$/F$_{11.2 \micron}$ between 0.5 and 1.4,
and follow the same trend as Galactic Disk and Magellanic Cloud PNe
(Bernard-Salas et al., in preparation).


\subsection{Dual Chemistry Nebulae}
\label{dual_chemistry}

{\it ISO} detected crystalline silicates and PAHs simultaneously in
[WR] PNe --- those PNe with H-poor and C-rich WR-type central stars
\citep{waters1998b}.  This dual chemistry is unusual in GDPNe
\citep{bernard-salas2005}. However, in our sample of GBPNe, six of the
eleven nebulae have dual chemistry, showing both crystalline silicates
and PAHs in their spectra. The fraction of [WR] PNe is significantly
larger in the Bulge than the Disk \citep{gorny2004}, and thus the
large fraction of PNe in the Bulge exhibiting dual chemistry makes
sense. Possible explanations for this dual chemistry include
\citep{littlemarenin1986, willems1986, waters1998b, cohen1999}: (1) a
thermal pulse recently ($\lesssim$ 1000 years ago) turned an O-rich
outflow into a C-rich one, and (2) the central star of the PN is in a
binary system and the silicate grains orbit the system in a disk that
existed long before the PN.

What explains how the majority of our GBPNe show dual chemistry?  The
explanation of a thermal pulse at the end of the AGB having suddenly
changed the chemical composition of the central star from O-rich to
C-rich within the last thousand years seems implausible because it is
unlikely that we would catch so many GBPNe in this short stage
\citep[e.g.][]{lloydevans1991}.  A growing body of evidence supports
the binary system with an old silicate disk explanation of dual
chemistry in PNe and late-type stars \citep{waters1998b, molster2001,
  matsuura2004}.  Taking one of these studies as an example,
\citet{matsuura2004} present mid-IR images of the post-AGB star IRAS
16279-4757 which shows both PAHs and crystalline silicates.  Their
images and model of this star imply that it has a C-rich bipolar
outflow with an inner low-density C-rich region surrounded by an outer
dense O-rich torus, indicating that mixed chemistry and morphology are
related; mixed chemistry may point to binary evolution.

Other evidence also suggests that many of our GBPNe probably have
binaries with silicate disks: (1) $\sim$40\% of compact PNe in the
Bulge have binary-induced morphologies \citep{zijlstra2007}; (2)
binary-induced novae are observed to be concentrated in the bulge of
the galaxy M31 \citep[e.g.][]{shafter2001,rosino1973}, and thus
perhaps in the Bulge of our galaxy as well; (3) asymmetric
(e.g. bipolar, quadrupolar) morphology is more common in PNe in high
metallicity environments than in low metallicity ones
\citep{stanghellini2003}; (4) the current study showing the similarity
of the mean GBPNe spectra to the mean disk spectra of
\citealt{molster2002} (\S \ref{crystalline_silicates_discussion}); and
(5) the silicates are crystalline and not amorphous, indicating that
they have been blasted over time and are likely stored in a disk
\citep{molster1999}. Thus it seems likely that the GBPNe in our sample
with dual chemistry have a binary at their center with a silicate disk
that formed long before the PN stage, while the PAHs reside in the PN
outflow itself, possibly shooting out along the poles.

\section{Conclusions}
\label{conclusion}

We extract the {\it Spitzer} IRS spectra of eleven PNe in the Bulge to
study their abundances and dust properties.   We conclude that:

{\bf (1)} The abundances of argon, neon, sulfur, and oxygen are
significantly lower in the PNe in the Bulge than the abundances for
the Bulge predicted by the abundance gradient in the Disk, consistent
with the idea that the Bulge and Disk evolved separately.

{\bf (2)} All of the spectra in our sample of PNe in the Bulge show
crystalline silicates, indicating that these crystalline silicates are
likely stored in disks, which would further imply that the progenitor
stars of these PNe evolved in binary systems.

{\bf (3)} Six of the eleven spectra of PNe in the Bulge in our sample
show PAHs in addition to the crystalline silicates.  This dual
chemistry also points toward binary evolution: the PAHs are in the
current PN outflow and the crystalline silicates reside in a old disk
created by binary interaction.

\acknowledgments This work is based in part on observations made with
the {\it Spitzer Space Telescope}, which is operated by the Jet
Propulsion Laboratory, California Institute of Technology under NASA
contract 1407. Support for this work was provided by NASA through
Contract Number 1257184 issued by JPL/Caltec. This research made use
of the SIMBAD database, operated at CDS, Strasbourg, France.




\begin{thebibliography}{}

\bibitem[{{Acker} {et~al.}(1992){Acker}, {Marcout}, {Ochsenbein}, {Stenholm},
  \& {Tylenda}}]{acker1992}
{Acker}, A., {Marcout}, J., {Ochsenbein}, F., {Stenholm}, B., \& {Tylenda}, R.
  1992, {Strasbourg - ESO catalogue of galactic planetary nebulae. Part 1; Part
  2} (Garching: European Southern Observatory, 1992)

\bibitem[{{Acker} {et~al.}(1991){Acker}, {Raytchev}, {Koeppen}, \&
  {Stenholm}}]{acker1991}
{Acker}, A., {Raytchev}, B., {Koeppen}, J., \& {Stenholm}, B. 1991, \aaps, 89,
  237

\bibitem[{{Allamandola} {et~al.}(1989){Allamandola}, {Tielens}, \&
  {Barker}}]{allamandola1989}
{Allamandola}, L.~J., {Tielens}, G.~G.~M., \& {Barker}, J.~R. 1989, \apjs, 71,
  733

\bibitem[{{Beaulieu} {et~al.}(1999){Beaulieu}, {Dopita}, \&
  {Freeman}}]{beaulieu1999}
{Beaulieu}, S.~F., {Dopita}, M.~A., \& {Freeman}, K.~C. 1999, \apj, 515, 610

\bibitem[{{Bensby} \& {Lundstr{\"o}m}(2001)}]{bensby2001}
{Bensby}, T. \& {Lundstr{\"o}m}, I. 2001, \aap, 374, 599

\bibitem[{{Bernard-Salas} {et~al.}(2008){Bernard-Salas}, {Pottasch},
  {Gutenkunst}, {Morris}, \& {Houck}}]{bernard-salas2008}
{Bernard-Salas}, J., {Pottasch}, S.~R., {Gutenkunst}, S., {Morris}, P.~W., \&
  {Houck}, J.~R. 2008, \apj, 672, 274

\bibitem[{{Bernard-Salas} \& {Tielens}(2005)}]{bernard-salas2005}
{Bernard-Salas}, J. \& {Tielens}, A.~G.~G.~M. 2005, \aap, 431, 523

\bibitem[{{Cahn} {et~al.}(1992){Cahn}, {Kaler}, \& {Stanghellini}}]{cahn1992}
{Cahn}, J.~H., {Kaler}, J.~B., \& {Stanghellini}, L. 1992, \aaps, 94, 399

\bibitem[{{Casassus} {et~al.}(2001){Casassus}, {Roche}, {Aitken}, \&
  {Smith}}]{casassus2001}
{Casassus}, S., {Roche}, P.~F., {Aitken}, D.~K., \& {Smith}, C.~H. 2001,
  \mnras, 327, 744

\bibitem[{{Ciardullo} {et~al.}(1999){Ciardullo}, {Bond}, {Sipior}, {Fullton},
  {Zhang}, \& {Schaefer}}]{ciardullo1999}
{Ciardullo}, R., {Bond}, H.~E., {Sipior}, M.~S., {Fullton}, L.~K., {Zhang},
  C.-Y., \& {Schaefer}, K.~G. 1999, \aj, 118, 488

\bibitem[{{Clegg} {et~al.}(1987){Clegg}, {Harrington}, {Barlow}, \&
  {Walsh}}]{clegg1987}
{Clegg}, R.~E.~S., {Harrington}, J.~P., {Barlow}, M.~J., \& {Walsh}, J.~R.
  1987, \apj, 314, 551

\bibitem[{{Cohen} {et~al.}(1999){Cohen}, {Barlow}, {Sylvester}, {Liu}, {Cox},
  {Lim}, {Schmitt}, \& {Speck}}]{cohen1999}
{Cohen}, M., {Barlow}, M.~J., {Sylvester}, R.~J., {Liu}, X.-W., {Cox}, P.,
  {Lim}, T., {Schmitt}, B., \& {Speck}, A.~K. 1999, \apjl, 513, L135

\bibitem[{{Condon} \& {Kaplan}(1998)}]{condon1998}
{Condon}, J.~J. \& {Kaplan}, D.~L. 1998, \apjs, 117, 361

\bibitem[{{Cuisinier} {et~al.}(2000){Cuisinier}, {Maciel}, {K{\"o}ppen},
  {Acker}, \& {Stenholm}}]{cuisinier2000}
{Cuisinier}, F., {Maciel}, W.~J., {K{\"o}ppen}, J., {Acker}, A., \& {Stenholm},
  B. 2000, \aap, 353, 543

\bibitem[{{Cunha} \& {Smith}(2006)}]{cunha2006}
{Cunha}, K. \& {Smith}, V.~V. 2006, \apj, 651, 491

\bibitem[{{Durand} {et~al.}(1998){Durand}, {Acker}, \& {Zijlstra}}]{durand1998}
{Durand}, S., {Acker}, A., \& {Zijlstra}, A. 1998, \aaps, 132, 13

\bibitem[{{Edgar} {et~al.}(2007){Edgar}, {Nordhaus}, {Blackman}, \&
  {Frank}}]{edgar2007}
{Edgar}, R.~G., {Nordhaus}, J., {Blackman}, E., \& {Frank}, A. 2007, ArXiv
  e-prints, 709

\bibitem[{{Eisenhauer} {et~al.}(2005)}]{eisenhauer2005} 
{Eisenhauer}, F., et~al. 2005, \apj, 628, 246

\bibitem[{{Escudero} {et~al.}(2004){Escudero}, {Costa}, \&
  {Maciel}}]{escudero2004}
{Escudero}, A.~V., {Costa}, R.~D.~D., \& {Maciel}, W.~J. 2004, \aap, 414, 211

\bibitem[{{Exter} {et~al.}(2004){Exter}, {Barlow}, \& {Walton}}]{exter2004}
{Exter}, K.~M., {Barlow}, M.~J., \& {Walton}, N.~A. 2004, \mnras, 349, 1291

\bibitem[{{Ferreras} {et~al.}(2003){Ferreras}, {Wyse}, \&
  {Silk}}]{ferreras2003}
{Ferreras}, I., {Wyse}, R.~F.~G., \& {Silk}, J. 2003, \mnras, 345, 1381

\bibitem[{{Fluks} {et~al.}(1994){Fluks}, {Plez}, {The}, {de Winter},
  {Westerlund}, \& {Steenman}}]{fluks1994}
{Fluks}, M.~A., {Plez}, B., {The}, P.~S., {de Winter}, D., {Westerlund}, B.~E.,
  \& {Steenman}, H.~C. 1994, \aaps, 105, 311

\bibitem[{{Fulbright} {et~al.}(2007){Fulbright}, {McWilliam}, \&
  {Rich}}]{fulbright2007}
{Fulbright}, J.~P., {McWilliam}, A., \& {Rich}, R.~M. 2007, \apj, 661, 1152

\bibitem[{{G{\'o}rny} {et~al.}(2004){G{\'o}rny}, {Stasi{\'n}ska}, {Escudero},
  \& {Costa}}]{gorny2004}
{G{\'o}rny}, S.~K., {Stasi{\'n}ska}, G., {Escudero}, A.~V., \& {Costa},
  R.~D.~D. 2004, \aap, 427, 231

\bibitem[{{Groenewegen} {et~al.}(2008){Groenewegen}, {Udalski}, \&
  {Bono}}]{groenewegen2008}
{Groenewegen}, M.~A.~T., {Udalski}, A., \& {Bono}, G. 2008, ArXiv e-prints, 801

\bibitem[{{Guiles} {et~al.}(2007){Guiles}, {Bernard-Salas}, {Pottasch}, \&
  {Roellig}}]{guiles2007}
{Guiles}, S., {Bernard-Salas}, J., {Pottasch}, S.~R., \& {Roellig}, T.~L. 2007,
  \apj, 660, 1282

\bibitem[{{Hanner} {et~al.}(1994){Hanner}, {Lynch}, \& {Russell}}]{hanner1994}
{Hanner}, M.~S., {Lynch}, D.~K., \& {Russell}, R.~W. 1994, \apj, 425, 274

\bibitem[{{Helou} \& {Walker}(1988)}]{helou1988}
{Helou}, G. \& {Walker}, D.~W., eds. 1988, {Infrared astronomical satellite
  (IRAS) catalogs and atlases. Volume 7: The small scale structure catalog},
  Vol.~7

\bibitem[{{Higdon et~al.}(2004)}]{higdon2004}
{Higdon}, S.~J.~U., et~al. 2004, \pasp, 116, 975

\bibitem[{{Houck et~al.}(2004)}]{houck2004}
{Houck}, J.~R., et~al. 2004, \apjs, 154, 18

\bibitem[{{Hummer} \& {Storey}(1987)}]{hummer1987}
{Hummer}, D.~G. \& {Storey}, P.~J. 1987, \mnras, 224, 801

\bibitem[{{Karakas}(2003)}]{karakas2003a}
{Karakas}, A.~I. 2003, {Asymptotic Giant Branch Stars: their influence on
  binary systems and the interstellar medium} (Thesis, Monash Univ. Melbourne.
  Available at http://www.mso.anu.edu.au/~akarakas/research.html)

\bibitem[{{Karakas} \& {Lattanzio}(2003)}]{karakas2003b}
{Karakas}, A.~I. \& {Lattanzio}, J.~C. 2003, Publications of the Astronomical
  Society of Australia, 20, 393

\bibitem[{{Knacke} {et~al.}(1993){Knacke}, {Fajardo-Acosta}, {Telesco},
  {Hackwell}, {Lynch}, \& {Russell}}]{knacke1993}
{Knacke}, R.~F., {Fajardo-Acosta}, S.~B., {Telesco}, C.~M., {Hackwell}, J.~A.,
  {Lynch}, D.~K., \& {Russell}, R.~W. 1993, \apj, 418, 440

\bibitem[{{Lecureur} {et~al.}(2007){Lecureur}, {Hill}, {Zoccali}, {Barbuy},
  {G{\'o}mez}, {Minniti}, {Ortolani}, \& {Renzini}}]{lecureur2007}
{Lecureur}, A., {Hill}, V., {Zoccali}, M., {Barbuy}, B., {G{\'o}mez}, A.,
  {Minniti}, D., {Ortolani}, S., \& {Renzini}, A. 2007, \aap, 465, 799

\bibitem[{{Little-Marenin}(1986)}]{littlemarenin1986}
{Little-Marenin}, I.~R. 1986, \apjl, 307, L15

\bibitem[{{Lloyd Evans}(1991)}]{lloydevans1991}
{Lloyd Evans}, T. 1991, \mnras, 249, 409

\bibitem[{{L{\'o}pez-Corredoira} {et~al.}(2000){L{\'o}pez-Corredoira},
  {Hammersley}, {Garz{\'o}n}, {Simonneau}, \& {Mahoney}}]{lopez-corredoira2000}
{L{\'o}pez-Corredoira}, M., {Hammersley}, P.~L., {Garz{\'o}n}, F., {Simonneau},
  E., \& {Mahoney}, T.~J. 2000, \mnras, 313, 392

\bibitem[{{Mart{\'{\i}}n-Hern{\'a}ndez} {et~al.}(2002)}]{martin-hernandez2002}
{Mart{\'{\i}}n-Hern{\'a}ndez}, N.~L., et~al. 2002, \aap, 381, 606

\bibitem[{{Matsuura} {et~al.}(2004)}]{matsuura2004}
{Matsuura}, M., et~al. 2004, \apj, 604, 791

\bibitem[{{Minniti} {et~al.}(1995){Minniti}, {Olszewski}, {Liebert}, {White},
  {Hill}, \& {Irwin}}]{minniti1995}
{Minniti}, D., {Olszewski}, E.~W., {Liebert}, J., {White}, S.~D.~M., {Hill},
  J.~M., \& {Irwin}, M.~J. 1995, \mnras, 277, 1293

\bibitem[{{Molster}(2000)}]{molster2000}
{Molster}, F.~J. 2000, PhD thesis, FNWI: Sterrenkundig Instituut Anton
  Pannekoek, Postbus 19268, 1000 GG Amsterdam, The Netherlands

\bibitem[{{Molster} {et~al.}(2002){Molster}, {Waters}, \&
  {Tielens}}]{molster2002}
{Molster}, F.~J., {Waters}, L.~B.~F.~M., \& {Tielens}, A.~G.~G.~M. 2002, \aap,
  382, 222

\bibitem[{{Molster} {et~al.}(2001){Molster}, {Yamamura}, {Waters}, {Nyman},
  {K{\"a}ufl}, {de Jong}, \& {Loup}}]{molster2001}
{Molster}, F.~J., {Yamamura}, I., {Waters}, L.~B.~F., {Nyman}, L.-{\AA}.,
  {K{\"a}ufl}, H.-U., {de Jong}, T., \& {Loup}, C. 2001, \aap, 366, 923

\bibitem[{{Molster} {et~al.}(1999)}]{molster1999}
{Molster}, F.~J., et~al. 1999, \nat, 401, 563

\bibitem[{{Peeters} {et~al.}(2002){Peeters}, {Hony}, {Van Kerckhoven},
  {Tielens}, {Allamandola}, {Hudgins}, \& {Bauschlicher}}]{peeters2002}
{Peeters}, E., {Hony}, S., {Van Kerckhoven}, C., {Tielens}, A.~G.~G.~M.,
  {Allamandola}, L.~J., {Hudgins}, D.~M., \& {Bauschlicher}, C.~W. 2002, \aap,
  390, 1089

\bibitem[{{Pottasch}(1984)}]{pottasch1984}
{Pottasch}, S.~R., ed. 1984, {Planetary nebulae - A study of late stages of
  stellar evolution}

\bibitem[{{Pottasch} \& {Beintema}(1999)}]{pottasch1999}
{Pottasch}, S.~R. \& {Beintema}, D.~A. 1999, \aap, 347, 975

\bibitem[{{Pottasch} \& {Bernard-Salas}(2006)}]{pottasch2006}
{Pottasch}, S.~R. \& {Bernard-Salas}, J. 2006, \aap, 457, 189

\bibitem[{{Pottasch} {et~al.}(2007){Pottasch}, {Bernard-Salas}, \&
  {Roellig}}]{pottasch2007}
{Pottasch}, S.~R., {Bernard-Salas}, J., \& {Roellig}, T.~L. 2007, \aap, 471,
  865

\bibitem[{{Ratag} {et~al.}(1997){Ratag}, {Pottasch}, {Dennefeld}, \&
  {Menzies}}]{ratag1997}
{Ratag}, M.~A., {Pottasch}, S.~R., {Dennefeld}, M., \& {Menzies}, J. 1997,
  \aaps, 126, 297

\bibitem[{{Ratag} {et~al.}(1992){Ratag}, {Pottasch}, {Dennefeld}, \&
  {Menzies}}]{ratag1992}
{Ratag}, M.~A., {Pottasch}, S.~R., {Dennefeld}, M., \& {Menzies}, J.~W. 1992,
  \aap, 255, 255

\bibitem[{{Reid}(1993)}]{reid1993}
{Reid}, M.~J. 1993, \araa, 31, 345

\bibitem[{{Rolleston} {et~al.}(2000){Rolleston}, {Smartt}, {Dufton}, \&
  {Ryans}}]{rolleston2000}
{Rolleston}, W.~R.~J., {Smartt}, S.~J., {Dufton}, P.~L., \& {Ryans}, R.~S.~I.
  2000, \aap, 363, 537

\bibitem[{{Rosino}(1973)}]{rosino1973}
{Rosino}, L. 1973, \aaps, 9, 347

\bibitem[{{Rubin} {et~al.}(1988){Rubin}, {Simpson}, {Erickson}, \&
  {Haas}}]{rubin1988}
{Rubin}, R.~H., {Simpson}, J.~P., {Erickson}, E.~F., \& {Haas}, M.~R. 1988,
  \apj, 327, 377

\bibitem[{{Ruffle} {et~al.}(2004){Ruffle}, {Zijlstra}, {Walsh}, {Gray},
  {Gesicki}, {Minniti}, \& {Comeron}}]{ruffle2004}
{Ruffle}, P.~M.~E., {Zijlstra}, A.~A., {Walsh}, J.~R., {Gray}, M.~D.,
  {Gesicki}, K., {Minniti}, D., \& {Comeron}, F. 2004, \mnras, 353, 796

\bibitem[{{Schutte} {et~al.}(1993){Schutte}, {Tielens}, \&
  {Allamandola}}]{schutte1993}
{Schutte}, W.~A., {Tielens}, A.~G.~G.~M., \& {Allamandola}, L.~J. 1993, \apj,
  415, 397

\bibitem[{{Shafter} \& {Irby}(2001)}]{shafter2001}
{Shafter}, A.~W. \& {Irby}, B.~K. 2001, \apj, 563, 749

\bibitem[{{Shaver} {et~al.}(1983){Shaver}, {McGee}, {Newton}, {Danks}, \&
  {Pottasch}}]{shaver1983}
{Shaver}, P.~A., {McGee}, R.~X., {Newton}, L.~M., {Danks}, A.~C., \&
  {Pottasch}, S.~R. 1983, \mnras, 204, 53

\bibitem[{{Simpson} {et~al.}(1995){Simpson}, {Colgan}, {Rubin}, {Erickson}, \&
  {Haas}}]{simpson1995}
{Simpson}, J.~P., {Colgan}, S.~W.~J., {Rubin}, R.~H., {Erickson}, E.~F., \&
  {Haas}, M.~R. 1995, \apj, 444, 721

\bibitem[{{Sloan} {et~al.}(2007)}]{sloan2007}
{Sloan}, G.~C., et~al. 2007, \apj, 664, 1144

\bibitem[{{Stanghellini} {et~al.}(2003){Stanghellini}, {Shaw}, {Balick},
  {Mutchler}, {Blades}, \& {Villaver}}]{stanghellini2003}
{Stanghellini}, L., {Shaw}, R.~A., {Balick}, B., {Mutchler}, M., {Blades},
  J.~C., \& {Villaver}, E. 2003, \apj, 596, 997

\bibitem[{{Tylenda} {et~al.}(1992){Tylenda}, {Acker}, {Stenholm}, \&
  {Koeppen}}]{tylenda1992}
{Tylenda}, R., {Acker}, A., {Stenholm}, B., \& {Koeppen}, J. 1992, \aaps, 95,
  337

\bibitem[{{van de Steene} \& {Zijlstra}(1995)}]{vandesteene1995}
{van de Steene}, G.~C. \& {Zijlstra}, A.~A. 1995, \aap, 293, 541

\bibitem[{{Walton} {et~al.}(1993){Walton}, {Barlow}, \& {Clegg}}]{walton1993a}
{Walton}, N.~A., {Barlow}, M.~J., \& {Clegg}, R.~E.~S. 1993, in IAU Symposium,
  Vol. 153, Galactic Bulges, ed. H.~{Dejonghe} \& H.~J. {Habing}, 337--+

\bibitem[{{Wang} \& {Liu}(2007)}]{wang2007}
{Wang}, W. \& {Liu}, X.~. 2007, ArXiv e-prints, 707

\bibitem[{{Waters} {et~al.}(1998)}]{waters1998b}
{Waters}, L.~B.~F.~M., et~al. 1998, \aap, 331, L61

\bibitem[{{Werner et~al.}(2004)}]{werner2004}
{Werner}, M.~W., et~al. 2004, \apjs, 154, 1

\bibitem[{{Willems} \& {de Jong}(1986)}]{willems1986}
{Willems}, F.~J. \& {de Jong}, T. 1986, \apjl, 309, L39

\bibitem[{{Zhang}(1995)}]{zhang1995}
{Zhang}, C.~Y. 1995, \apjs, 98, 659

\bibitem[{{Zijlstra}(2007)}]{zijlstra2007}
{Zijlstra}, A.~A. 2007, Baltic Astronomy, 16, 79

\end{thebibliography}

\end{document}